\documentclass[a4paper,11pt]{article}
\usepackage[a4paper]{geometry}

\usepackage{adjustbox}
\usepackage{amsmath}
\usepackage{amssymb}
\usepackage[T1]{fontenc}        
\usepackage{lmodern}    
\usepackage[utf8]{inputenc}     
\usepackage{ifthen}

\usepackage{graphicx}
\graphicspath{ {./images/}} 
\usepackage{adjustbox}

\begin{document}
\title{Approximate $SU(5)$, Fine Structure Constants }
\author{
H.B.Nielsen, Niels Bohr Institute}

\maketitle

\abstract{We fit the three finestructure constants of the Standard Model with
  three,
  in first approximation theoretically
  estimable
  parameters, 1) a ``unified
  scale'',turning out {\em not} equal  to the Planck scale and thus 
  only estimable by a very speculative story, 2) a ``number of layers'' being
  a priori the number of families, and 3) a unified coupling related to a
  critical
  coupling on a lattice. So formally we postdict the three fine structure
  constants!
  In the philosophy of our model there is a physically
  lattice theory with
  link variables taking values in a (or in the various) ``small''
  representations of the
  Standard Model {\bf Group}. We argue for that these representations function
  in first approximation as were the theory a genuine $SU(5)$ theory. Next we
  take into account fluctuation of the gauge fields in the lattice and obtain a
  correction to the a priori $SU(5)$ approximation, because of course the
  link fluctuations not corresponding any Standard model Lie algebra, but
  only to the SU(5), do not exist.
  The model is a development of our old anti-grand-unification model having as
  its genuine gauge group, close to fundamental scale, a cross product of the
  standard model group $S(U(3)\times U(2))$ with itself, there being  one
  Cartesian product factor for each family. 
  In these old works we included the hypotesis of ``multiple point criticallity
  principle'' which here effectively means the coupling constants be critical
  on the lattice. Counted relative to the Higgs scale we suggest the in our
  sense ``unified scale'' (where the deviations between the inverse fine
  structure constants deviate by quantum fluctuations being only from
  standard model groups, not SU(5)) makes up the 2/3 th power of the Planck
  scale relative to the Higgs scale, or better the topquarkmass scale..}
\begin{center}
  {\large Keywords}
  \end{center}

{grand unfication SU(5); lattice theory; running couplings; anti-GUT;
  Standard Model {\bf Group};critical coupling; energy scales;
  fine structure constant{\bf s}; Planck scale } 

\vspace{1mm}

\section{Introduction}

We\cite{RDrel1,RDrel2,RDrel3,RDrel4,Bled8,Don93,LRN,flipped, Takanishi} and
others\cite{Volovik,
  Picek,Ryzhikh, LR}
have long - long time ago - worked on fitting the fine structure constants -
especially the
non-abelian ones - in a model based on the main assumptions:
\begin{itemize}
\item{\bf Critical Couplings at Fundamental Scale} Preferably the gauge
  couplings should be at some multiple critical point for a lattice theory
   at the``fundamental scale'' .
  And it
  was in the spirit of that model, that there indeed would exist a lattice
  theory
  in Nature.
\item{\bf AntiGUT} The gauge group  was at the ``fundamental scale'' the
  Cartesian product $G\times G \times ...\times G$ of the same
  group $G$ with itself, one time for every family of fermions.
  \end{itemize}
but mainly the Abelian coupling of $U(1)$ was not so well predicted contrary
to the non-abelian ones (the attempt by Don Bennett and myself \cite{Don137}
got good numbers, but the theory is a bit complicated). Further Laperashvili
and Das and Ryzhikh \cite{LD,LRSU5,LR} have even united this type of model with
grand unification with $SU(5)$ \cite{LD,LRSU5}. They used also supersymmetry
in their
picture.

Now it is the point of the present article to also make such a combination
of $SU(5)$ GUT\cite{GG, GG2
} and the A(nti)GUT theory(AGUT= ``anti grand unification theory''
meaning the type of theory with a cross product of several copies of the
standard model group, e.g. one cross product factor for each family
of fermions) just mentioned, but {\bf without
  SUSY}. Rather we shall here seek an $SU(5)$-like ``unification''
{\bf without taking
  the $SU(5)$ theory as really true}, but rather taking the $SU(5)$ as
  an {\bf approximate symmetry}  appearing, because of the link variables have
  a form reminding of $SU(5)$.
  In fact one possible argumentation is to  assume, that the link variables
  are constructed as matrices (with dynamical matrix element with somewhat
  restricted movability)
  for a most simple and smallest faithfull representation (a sort of
  principle \cite{sr, srKorfu14,srcim4}
  of smallest link-representation).
  Another similar argumentation is to use our earlier work
  \cite{sr, srKorfu14,srcim4}
  telling,
  that one can define a concept of ``small representation'' so that the
  standard model {\bf group}\cite{OR}\footnote{O Raifeartaigh points
    out that by choosing the {\bf group} among the set of groups,
    with the given Lie algebra, which is ``smallest'' and thus
    have the fewest representations, but still has the representations
    used by the fermions and the Higgs(es), one can claim, that one
    selected {\bf the} {\bf gauge group} for the used theory with its
    fermions etc. So a sense can be given in this way to
    {\bf the Standard Model Group} and it turns out to be
    $S(U(2)\times U(3))$ meaning the group of $5\times 5$ matrices
    composed along (and around)  the diagonal a $U(2)$ and a $U(3)$ and then
    impose the condition - symbolized by the ``$S$'' -
    of the determinat of the whole $5\times 5$ matrix being
    $det =1$ gets selected as having smallest faithfull
  representation among all groups.}  This would, taken seriosly, tell, that
  it is important, that the group chosen by Nature should have small
  representation, and that makes it natural that the link degrees of
  freedom corresponds to a ``small'' faithfull representation of the
  standard model group. Then it turns out, that a typical such small
  representation is the one obtained by starting from the ${\bf 5 }$-plet
  of $SU(5)$ and restrict to the Standard Model Group as contained in $SU(5)$.
  Really 
   the standard model group
   $S(U(2)\times U(3))$ is even in the notation  as used here an obvious
   subgroup of just $SU(5)$, the notation of which- the $5\times 5$ matrices -
   is used to write it.

   In the game we proposed\cite{sr,srKorfu14, srcim4} to specify the
   Standard model group as a {\bf group},
   it turns out that  a cross product of several isomorphic groups gets the same
   ``points''(in the game of our reference \cite{sr,srKorfu14, srcim4}
   so that the AGUT model believed in the article is on a shared first place
   with the single Standard Model {\bf group}) as the group itself, so a
   group $G_{SMG}\times G_{SMG}\times ...
   G_{SMG}$ would be equally favoured by the our game.

   .
   In any case the idea is, that the link variables are in terms of the
   fundamental physics, that is imagined to be behind, represented by
   variables
  like in some ``small'' representation \cite{sr} of the standard model group,
  and that then this representation happens to be / naturally is effectively
  an $SU(5)$-representation. This means that the link variables can formally
  be interpreted as $SU(5)$ variables; but  {\bf in reality they are not}.
  (i.e. there is {\em no} $SU(5)$ symmetry for turning around the
  matrix elements
  in link {\bf 5}-plet, {\bf only under the Standard Model subgroup}.)
  There is {\bf no true $SU(5)$ theory in our model}! But we can describe
  the model
  in terminology of an $SU(5)$-theory, which is broken fundamentally. It is
  not broken by Higgs mechanism as in the usual SU(5)-theories (a priori
  at least), but
  other gauge fields than the ones in the standard model subgroup do not
  exist (in the first place). There are only gauge fields corresponding to
  the degrees of freedom
  in the standard model groups - one set for each family, however, -.
  (So you must imagine either, that we really have the gauge group
  $G_{SMG}\times G_{SMG} \times \cdots \times G_{SMG}$ with as many standrad model
  group factors as there are families of fermions, 3, or you imagine there to
  be three layers of a usual lattice, so that we have three links, where you
  usually have only one.)

  In the very crudest approximaton for a lattice action - linear in the
  trace of the representation matrix, the similarity to the $SU(5)$
    matrix theory is so great, that the coupling constant ratios at the
    fundamental lattice theory
    in the first approximation become
    just as in the GUT $SU(5)$ unification scale. However, when it
    now comes to perturbative corrections due to the fluctuation
    of the lattice theory degrees of freedom, it becomes important
    that the degrees of freedom present in $SU(5)$ theory, but not in the
    Standard model, are missing, and therefore cannot fluctuate. So the
    quantum corrections from the fluctuation of these - in standard model
    not present - degrees of freedom are lacking, and thus makes the effective
    couplings observed in the continuum limit get different values from what
    they
    would have gotten in a true $SU(5)$ theory. Being quantum corrections
    one would usually treat them perturbatively and expect them to be small.
    If this is indeed the case, then the usual $SU(5)$ predictions
    will be {\bf approximate}! We can say that it is the main point of
    present article to calculate this deviation from the exact $SU(5)$
    predictions to the usual picture of unifying gauge couplings. Thus the
    Standard Model (inverse) fine structure constants do not
    truly unify (at a unification scale, but we shall talk in the present
    paper about an ``our unified scale'', which is the scale at which
    there is unification except for our (quantum) corrections, and that we
    call $\mu_U$), but we calculate the
    degree of lack of unification, and even
    make {\bf prediction of the numerical value of the deviation.}

\subsection{Some Previous attempts}
\label{threferee1}
Since it has long been wellknown that using minal SU(5) there is no working
unifying scale in the sense that we see on e.g. fig 1 that the three lines
representing the runnings of the three Standard Model Lie groups do not cross
just in a point as SU(5) GUT would say. The two very popular ways of seeking
to solve this problem are :
\begin{itemize}
\item{Super symmetry} By introducing supersymmetry one has to have a scale
for breaking this susy and thus of course a new parameter with which to
adjust the crossing of the three lines. It happens and this is then a
success for susy that this susy-breaking scalebeing very close where we have
the data and thus to the place from which susy is possible. We namely expect
from the hierarchy problem which susy can rescue that the breaking scale
shall be very in energy. So susy like our model has a theoretical story
about the value of parameter introduced to fit the SU(5). We have
instead that our quantum correction shall be just the number of families times
bigger than calculated with a simple lattice.

\item{Bigger Groups containing SU(5)}

If you have a group containing SU(5) as a subgroup such as SO(10) there are
possibilities for having the three standard model groups so to speak be packed
into bigger simple groups at different ranges of scales and thus also
get a way to achieve an extra parameter and thus solve the problem.

In our philosophy we start from the {\bf group} of the Standard Model which
is $S(U(2)\times U(3))$, which means we already have the favoured group fixed,
and then argue that the ``smallest'' / essentially simplest representation to
use to construct the plaquette variable action contribtuion {\bf happens
to be an $SU(5)$ representation}. That is to say we claim that taking the
Standard Model very seriously there is pointing to the group $SU(5)$ proper,
in a way we do not have for the alternative unification groups.

This is of course a very weak argument in favour of just SU(5) only, and
thus in theories, that have some prediction for the energy scales of the
various symmetry breakings, a prediction of the minimal SU(5)-breaking
in the fine structure constants would be of an anlogous strength as our
somewhat ad hoc factor 3 on the quantum correction fromthe lattice.

\end{itemize}

In general one should bring it as a propaganda for our model, that apart
from the lattice - which should at the end be assumed to be fluctuating
in link size from place to place in Minkowski space as quantum fluctuations -,
a lattice that is somehow needed to avoid divergencies, we keep to just the
Standard Model and nothing else, at least if you ignore our story about the
critical coupling, which is only important for settling the unified fine
structure constant. In other words {\bf our model claims to be more minimal}
than the
just mentioned competing scenarious susy and grand unification with
larger than SU(5) groups,or even with true SU(5) full symmetry.

    \subsection{Character of Our Prediction(s)}
\begin{figure}
  \includegraphics[scale=1.3]{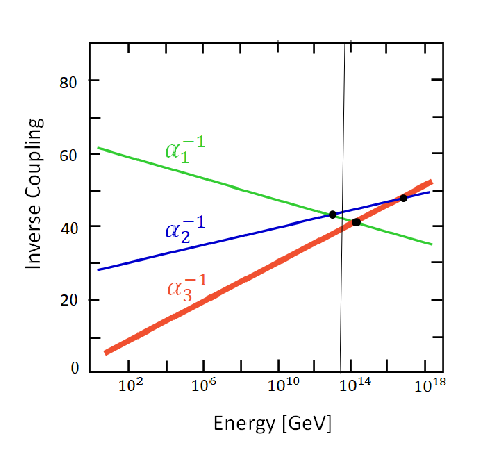}
  \caption{This is the usual graph representing the three Standard Model
    inverse fine structure constants with the $\alpha_1^{-1}$ being in the
    notation suitable for $SU(5)$, meaning it is 3/5 times the natural
    normalization, $\alpha_{1\; SU(5)}^{-1}$ =$ 3/5 *\alpha_{1\;SM}^{-1}$ =$
    3/5*\alpha_{EM}^{-1} cos^2\Theta_W$. The vertical thin line at the energy
    scale
    $\mu_U = 5*10^{13} GeV$ points out ``our unified scale'', which is as
    can be seen not really unifying the couplings, but rather is the scale
    where the ratio of the two independent differences, $\alpha_2^{-1}
    -\alpha_{1\; SU(5)}^{-1}$ and $\alpha_{1\; SU(5)}^{-1} - \alpha_3^{-1}$ have
    just the ratio 2/3 as our model predict at the `` our unification scale''.
    One may note that this ``our unified scale'' is actually very close to,
    where the three inverse couplings are nearest to each other, and in that
    sense an ``approximate'' unification scale.\label{fig1}}
\end{figure}

\begin{figure}[h]
  \includegraphics[scale=1.3]{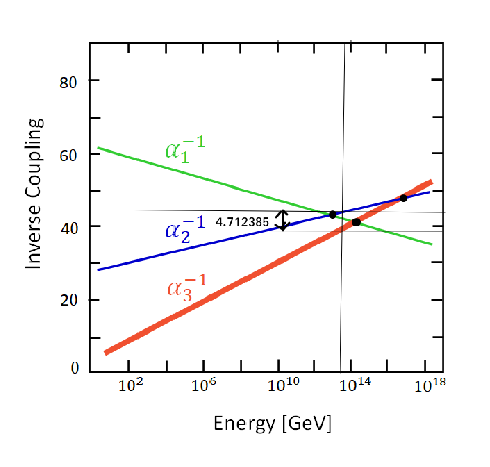}
  \caption{Same as figure 1, but now with our prediction inserted, marked as
    the number $4.712385 = 3*\pi/2$, which is predicted to be at the ``our
    unified scale''the difference $1/\alpha_2 -1/\alpha_3$. Our prediction
    is, that just at horizontal thin black line at $5*10^{13} GeV$ corresponding
    to the scale $\mu_U$, given by our fitted to the green line crossing point
    dividing the region between the blue and the red in the ratio 2 to 3, we
    shall have
    the difference in ordinate between the red and the blue crossing points
  with the vertical black being $3\pi/2$. \label{fig2}}
\end{figure}

    The main point of the present article (recently further developped in
    \cite{RSR, BledRSR}) is really to predict the deviation
    from exact $SU(5)$ GUT at a certain scale $\mu_U$ at which we calculate
    the corrections to the exact $SU(5)$ inverse fine structure constants
    in the standard model as due to quantum fluctuations in the lattice theory
    assumed to be really physically existent at some scale. Since we predict
    the absolute values of the differences between the inverse finestructure
    constants at the scale, we have at this scale two numerical predictions,
    and thus we can afford to use one of these predictions at the fundamental
    scale to fit the scale, and we shall still have one predition left. For
    instance we can use the prediction at the scale, at which the ratio of the
    difference $1/\alpha_2(\mu_U) - 1/\alpha_{1\; SU(5)}(\mu_U)$ to the other
    difference
    $1/\alpha_{1\; SU(5)}(\mu_U) - 1/\alpha_3(\mu_U)$ shall be 2 to 3
    (as our calculation implies). This is
    illustrated
    on figure \ref{fig1}, and one shall remark, that the three crossings of the
    inverse fine structure constant with the vertical black line on the
    figure at the scale about $5.*10^{13} GeV$ has been fitted, so that the
    three crossings lie with the ratio 2 : 3 of the two intervals. The $U(1)$
    inverse fine structure
    constant passes in between the $SU(2)$, above it with a piece that
    is proportional to 2, and the $SU(3)$ line, then below it with a distance
     proportional to 3. But having fixed the scale $\mu_U$ this way it
    is still a very nontrivial prediction that e.g. the absolute difference
    between the $SU(2)$-crossing and the $SU(3)$-crossing is just
    $3\pi/2 = 4.712385$. This is illustrated on figure \ref{fig2}.


\subsection{Our Rather Simple Fitting Formulas }

\subsubsection{The Quantum Corrections Breaking the Approximate $SU(5)$}

Our formulas for fitting the three inverse finestructure constants in the
Standard Model in the for $SU(5)$ adjusted notation, wherein one uses
$1/\alpha_{1 \: SU(5)} = 3/5 * 1/\alpha_{1 \;  SM}= 3/5 *cos^2\Theta_W*1/\alpha_{EM}$
are rather simple, and concerns of course the three Standard Model fine
structure by renormalization group transformed to a certain scale $\mu_U$, which
is our replacement for the unification scale (because there is of course
as we know no unification scale proper unless one involves susy or something
else extra). The choice of the scale $\mu_U$ is only indirectly determined in
our model, and is essentially just a fitting  parameter, although we in section
\ref{scale} shall speculatively relate $\mu_U$ to the Planck energy scale
$E_{Pl}$ by
a crude relation $\frac{\ln(\frac{E_{Pl}}{m_t})}{\ln(\frac{\mu_U}{m_t})}
\approx 3/2$. Then at this scale - to be
fitted -
\begin{eqnarray}
  \frac{1}{\alpha_{1 \; SU(5)}(\mu_U)}&=& \frac{1}{\alpha_{5\; uncor.}}- 11/5 *q\\
  \frac{1}{\alpha_{2}}(\mu_U)&=& \frac{1}{\alpha_{5\; uncor.}}- 9/5 *q\\
    \frac{1}{\alpha_{3}(\mu_U)}&=& \frac{1}{\alpha_{5\; uncor.}}- 14/5 *q,
\end{eqnarray}
where the one parameter $\frac{1}{\alpha_{5\; uncor.}}$, which could also give
other names like
\begin{eqnarray}
  \frac{1}{\alpha_{5 \; bare}} &=& \frac{1}{\alpha_{5 \;classical}}=
  \frac{1}{\alpha_{5 \; uncor.}},
\end{eqnarray}
is our replacement for the unified inverse $SU(5)$ fine structure constant.
The symbols, which we propose $uncor.$ = $bare$ = $classical$ are to tell
that this coefficient in the action functioning as the $ SU(5)$ inverse
coupling is without the quantum fluctuation couplings, i.e. it is
uncorrected (= uncor.) or ``bare''. We could also define a corrected one
\begin{eqnarray}
  \frac{1}{\alpha_{5 \; cor.}} &=& \frac{1}{\alpha_{5\; uncor.}}- 24/5*q.
\end{eqnarray}

  The other parameter $q$ we believe to calculate in our model with
  its 3 families of fermions and in a Wilson lattice in a lowest order
  approximation:
  \begin{eqnarray}
    q&=& ``\# families''*\pi/2 = 3*\pi/2 = 4.712385.
    \end{eqnarray}
  Using this notation we could equally well use the formulation
  \begin{eqnarray}
    \frac{1}{\alpha_{1\; SU(5)}(\mu_U)}&=& \frac{1}{\alpha_{5\; cor.}}+ 13/5 *q\\
       \frac{1}{\alpha_{2}(\mu_U)}&=& \frac{1}{\alpha_{5\; cor.}}+ 3 *q\\
       \frac{1}{\alpha_{3}(\mu_U)}&=& \frac{1}{\alpha_{5\; cor.}}+ 2 *q.
    \end{eqnarray}
  Here in fact the quatity $\frac{1}{\alpha_{5 \; cor.}(\mu_U)}$ is the in
  the analogous
  way to our treatment of the Standard Model inverse fine structure constants 
  formally corrected  $SU(5)$- inverse coupling to an effective  one at the
  our unified scale $\mu_U$, but of
  course,
  since there is no $SU(5)$, this is not so important, and rather formal only.
  \subsubsection{The Critical Coupling}

  The requirement of the gauge couplings at the fundamental scale being
  just on the borderline on one or preferably more phase transitions,
  that are welcome to be lattice artifacts, was the basic ingredient in
  the wroks, of which the present one is a development\cite{RDrel1,RDrel2,
    RDrel3,RDrel4,Bled8,Don93,LRN,flipped, Takanishi}. In the present
  work with its approximate $SU(5)$ it may seem natural to require
  the $SU(5)$ coupling
  being just on the phase border for the pseudo-unified $SU(5)$ coupling
  as represented by $\frac{1}{\alpha_{5\; uncor.}}$. In principle the
  critical coupling depends on the lattice details, and it has to be calculated
  by lattice computer calculations, but here we have for a beginning just
  taken an approximate formula for the critical coupling out of our
  earlier works\cite{LRNwf}.

  \subsubsection{The ``Unified Scale'' from in Lattice Constant Fluctuating
    ``Lattice''}

  The fact, that has always been a bit embarrasing for GUT theories of e.g.
  $SU(5)$, namely that the unified scale turns out appreciably smaller in energy
  than the Planck scale, is also embarrasing in our theory, and for rescuing
  it against this problem, we propose the speculation of a strongly fluctuating
  lattice. It should fluctuate  in the size of the lattice constant, and we
  should imagine, that in various places and moments the lattice is more or less
  fine. We shall below see, that this kind of fluctuations can be used as an
  excuse for the effective scale for gravity, the Planck energy scale,
  and that for the Standard Model, the ``our'' grand unified scale
  (which is a replacement for the GUT scale) can deviate from each
  other violently. The parameter giving the our unified scale
  $\mu_U$, namely the logarithm of it relative to the weak scale $M_Z$, namely
  $\ln(\frac{\mu_U}{M_Z})$ (or may be use better $m_t$ instead of $M_Z$), is
  according to our speculation given in terms
  of the Planck scale, which thus is a needed input to obtain all three
  parameters to give the three fine structure constants.

  \subsubsection{Resume of the Fitting}

  The three parameters, with which we fit the three Standard Model
  fine structure constants come
  in our present work from rather different speculations, which  though all
  should be sufficiently compatible, that we can have them in the same
  model. Here we announce, in the below table, the success of our model:
  
  \begin{adjustbox}{max width =\textwidth} 
  \begin{tabular}{|c|c|c|c|c|c|}
    \hline
    Parameter& Formula & From $\alpha$'s & Theory & Deviation&Section\\
    \hline
    q &q=$1/\alpha_2(\mu_U) -1/\alpha_3(\mu_U)$ &4.618201&  4.712385
    &-0.094$\pm$0.05  &\ref{ps3},
    \ref{dbausu5}\\
    \hline
    $1/\alpha_{5 \; uncor.}(\mu_U)$ &see above& 51.705&45.927&5.778$\pm$ 3.5
    &\ref{ps10}\\
    \hline
    $\ln(\frac{\mu_U}{M_Z})$ &$\ln(\frac{\mu_U}{m_t}) =\frac{2}{3}*
    \ln(\frac{E_{Pl\; red}}{M_t})$  & 26.43 &24.76 &1.67$\pm$ 1 
    &\ref{scale}\\
    &&&&or 0.02&\\
    \hline
\end{tabular}
  \end{adjustbox}
  
  In the third parameter line we put a somewhat by hand taken
  uncertainty for the theoretical value, because the scales being
  diveded,the Planck scale over the scale of the three families ending
  at low energy taken as the $M_Z$ scale or better top-mass $m_t$, is a ratio
  of rather illdefined
  concepts of scales and thus at least give an uncertainty of one unit
  in the natural logarithm.

  Depending on how many of the stories behind the ``theory''  of these
  parameters the reader might buy as trustable the reader can decide with how
  many parameters, we fit the three standard model (inverse) fine structure
  constants. In fact the ``theories'' for the three different parameters are
  rather independent of each other, so that a selections that some are wrong and
  some are right would not at all be excluded.
  
\subsection{Plan of Article}

In the following section \ref{model} we describe our, the assumption of
lattice for the Standard Model {\bf Group}, which means that it is important
also, what the global structure of this group is, and not only the Lie algebra.
According to O'Raifaighty \cite{OR} the global structure of group
is connected with the system of allowed representations, and one thus can
consider the system of repsentations for the fermions in the standard
model as a strong indication for the special gauge {\bf group} $S(U(2)\times
U(3))$.

In section \ref{ps3} we perform calculations of the quantum correctons
meaning calculating  zero-point fluctuations in plaquette variables, Taylor
expanding the partition function and developping a table for the contributions
of the zero point fluctuations on the continuum/effective (inverse) fine
structure constants. Strictly speaking our correction depends on the
type of lattice used, although we hope that it will be very little dependent.
In the
section \ref{Wil} we at least mention the Wilson lattice action, which so to
speak is the lattice we have used. 
In section \ref{ps6} we compared to an old similar
quantum fluctuation  which we, Don Bennett and the present author, made many
years ago in the similar model. Also we look for checking of  our too  crude
estimation
for what in lattice calculations is called tadpole improvements
\cite{Niyazi}, but actually is the quantum fluctuations, we consider 
being the main mechanism breaking the approximate $SU(5)$, that appeared so to
speak by accident, because the representation matrix in the links happened
to have also $SU(5)$ symmetry, before some motions of it are restricted not
to occur.

The fitting of the data - the experimentally determined fine structure
constants in the Standard Model - comes in section \ref{ps7}, where we first
determine by the requirement of the ratio of the difference between running
couplings being as we predict the scale, that must be the fundamental scale
in our model $\mu_U$. It is what we can call ``our unification scale'' $\mu_U$,
but really of course there is no true unification, since our $SU(5)$ is only
approximate. Next we compare, if the seperations at this scale is what we
predict. At the end of the section we do it oppositely, as a check.

In section \ref{ps10} we look at, if the coupling, say the approximate GUT one,
is the critical coupling. In the works, which led up to this one, this having
critical couplings were the crucial point\cite{RDrel1,RDrel2,RDrel3,RDrel4,
  Bled8,Don93,LRN,flipped, Takanishi}.
We shall in general postpone second order calculation, but we should mention
that a second order calculation is called for, see section \ref{ps11}, and presumably not
exceedingly
hard.

In section \ref{scale} we discuss the most speculative one among the parmeters
in our model, which should be obtainable in an other way than just by the
fitting fine-structure constants, namely the ``scale of unification'' $\mu_U$.
Although it is probably the most chocking result, if one would
believe our model, that {\bf the ``fundamental scale $\mu_U$ is not
  the Planck scale}, then we shall present a speculative story on in
size fluctuating lattice, that shall suggest a relation between
the ``fundamental scale'' in our model and the Planck one (allowing them
to deviate in order of magnitude).

Finally in section \ref{ps12} we conclude, but also include some thoughts
about the problems or suggestion for a quantum gravity, if we take the present
work so serious, that we must claim that the fundamental scale for the Standard
Model is the unification one, for our approximate $SU(5)$ GUT, even a bit
low in energy compared to the usual unified scale. A lattice, which fluctuates
even in scale in some background of a manifold or a projective space. If one
could have the lattice imbedded in the continuum space with some symmetry
including scalings, there might be a chanse of having a different
way to average over fluctuations in the lattice constant size
(i. e. coping with a fluctuating ``fundamental
scale'') for the fine structure constants gauge theories and for the
gravity. Such different averaging can seperate the different scales to be
observed for the two groups of forces,the Standard Model ones , and gravity. 

\section{Our Model}
\label{model}

    Our concrete model is, that we have in Nature a fundamental lattice with
    an energy scale $\mu_U$ crorresponding to the lattice constant $1/\mu_U$
    (with $c=\hbar=1$), the lattice being the Wilson one, say.
    This lattice is ``tripled
    up'' in the
    sense, that there is really one Wilson lattice for each family of fermions.
    Calling the number of families $N_{gen}=3$ one can think of it as the
    genuine  group being not the Standard Model Group itself $SMG$,  but its
    third
    power $SMG\times SMG \times SMG$, the true gauge group in our model
    \begin{eqnarray}
      G_{full}&=& SMG\times SMG \times SMG\\
      \hbox{where } SMG &=& S(U(2)\times U(3))\\
      &=&({\bf R}\times SU(2) \times SU(3) )/{\bf Z}_{app}\\
      \hbox{where } {\bf Z}_{app}&=& \left \{ (r, U_2, U_3) |\exists
      n \in {\bf Z}
      [(r,U_2,U_3) =\left ( 2\pi,-{\bf 1}, \exp(i2\pi/3) {\bf 1}\right )^n]
      \right \}\nonumber
    \end{eqnarray}

     The symbol ${\bf Z}_{app}$ denotes a discrete subgroup of the center
of the covering group ${\bf R}\times SU(2) \times SU(3)$ which is
determined by the Lie algebra of the Standard Model. The crucial requirement
is that the elements is this subgroup of the center shall be represented
trivially in all the representation for the various fields/particels in the
Standard Model. This requirement leads to a rule ensuring in fact that leptons
must have integer electric charges and and that the quarks must have charges
1/3 modulo 1. This means that some of the victories usually ascribed to SU(5)
grand unification here is considered built into the global group.
    
    Alternatively one might think of a model like this as there being
    three usual lattices lying parallel to each other (seperated in an extra
    dimension, say), It could therefore be tempting to call them
    ``layers'' of lattices.

    In any case we imagine, that somehow or another the $G_{full}$ is broken
    down to its diagonal subgroup, which is (isomorphic to) the standard Model
    group $SMG$. In fact this diagonal  subgroup is defined as
    \begin{eqnarray}
      SMG_{diag} &=& \hbox{the subgroup of } G_{full}\hbox{of elements of form
        $(g,g,g)$}\nonumber\\
      SMG_{diag}&=& \{(g,g,g) \in G_{full} = SMG\times SMG \times SMG |
      g\in G_{SMG}\}\nonumber. 
    \end{eqnarray}
    (we tend to use both notations $SMG$ and $G_{SMG}$ for the same,so
    simply $SMG = G_{SMG}=S(U(2)\times U(3))$).
    This breaking down of $G_{full}$ to the diagonal $SMG_{diag}$ can easily be
    imagined to come about by a little bit of mixing up the different
    layers locally all over. ( ``confusion'' \cite{Polonica, confusionetal,
      Mizrachi}).
    In the section \ref{scale} we shall speculate a bit more complicated about
    the lattice structure, because we shall propose that there is even at the
    lattice scale diffeomorphism symmetry or at least some symmetry, like
    the symmetry of a projective space time containing (local or global)
    scalings. This then means that we imagine the lattice to fluctuate
    in both size and position, so that even if it is Wilson type very locally,
    it varies in both orientation and size of the lattice constant
    very strongly from place to place. If it is so, and it might be unrealistic
    to imagine that it is not fluctuating, if we shall have a so  usual
    gravity theory with its reparametrisation fluctuating (as one should
    imagine the gauge of any gauge theory to really fluctuate \cite{FNN}),
    e.g. the ``fundamental scale''  $\mu_U$ we calculate below by fitting,
    must be at the end considered an average value of the ``fundamental scale''
    while the local fundamental scale fluctuates.

    But apart from this story of connecting our model to gravity, the
    fluctuations might be ignored, and a lattice with fixed lattice
    constant of order $a\sim 1/\mu_U$  would be o.k. (But remember:
    we fit ``the our unification scale'' like the one in usual
    exact $SU(5)$ to be appeciably lower in energy than the Planckscale.)
    
    \subsection{The ``Small'' Representations Used in the Links and Plaquettes}
    The crucial special assumtption for this article is to assume, that the
    degrees of freedom of the lattice-links representing the element of the
    standard model group SMG is the matrix elements of a matrix representation
    of this SMG on a minimal faithfull representation. It is then assumed that
    these matrix elements are restricted to only (be able to) move quite
    freely along the
    image of the SMG into the ``small'' representation used, while motion
    in other directions is strongly restricted (perhaps by very high
    potentials)
    but at least we shall ignore them, if there is any fluctuations, except
    along
    the standard model group, so to speak. The idea of thinking of such an
    imbedding is to note, that in such an imbedding we have a way of thinking
    of an SU(5) representation too, because the ``small'' reprensentation,
    we have in mind, is the one, that is the ${\bf 5}$ plet reprentation
    of $SU(5)$. It is of course also a representation of the $SMG\subset
    SU(5)$. Now a really crucial point is, that we imagine, that once the
    $SMG$ has been represented this $SU(5)$ simulating way, it tends to inherit
    an $SU(5)$ symmetry, even though our model has {\bf no true $SU(5)$
      symmetry postulated}. It is only, that it seems a bit similar in its
    simplest reprentaion. A bit more concretely we may say: we use, that the
    smoothness assumed also for the Lagrangian density as function of the
    plaquette- variables - which are also postulated to be formulated in
    $5\times 5$ matrices - is a smoothness defined from the $5\times 5$
    matrices. When we then Taylor expand and from that look for the form
    of the plaquette action, we come to the trace of the $5\times 5$ matrix
    just as in the usual $SU(5)$ theory. By this we have thus ``sneaked in''
    an {\bf approximate} $SU(5)$ symmetry. This is really the crux of matter
    of our model: The $SU(5)$ symmetry is {\bf not a symmetry imposed on Nature}
    but rather an approximate symmetry of the way, we suggested to be the most
    natural way to represent the link and plaquette degrees of freedom
    for a model, that basically is only symmetric under the standard
    model group SMG. Thus there is of course already in our picture built
    in a breaking of the $SU(5)$ symmetry. Most importantly the {\bf degrees of
    freedom from the components in the $SU(5)$ theory fields not also in the
    Standard Model Group SMG, are lacking.}

    For us this then means, that there are no quantum fluctuations in the
    plaquette or link variables corresponding to these lacking degrees of
    freedom. The concern of the present article is to evaluate, how
    these lacking modes lead to lakcing some quantum corrections for the fine
    structure constants, and these corrections from the lacking modes
    of oscillations are not quite equally big for the three different
    Standard model gauge couplings. This is then according to us the
    reason for breaking in these couplings of the - of course fundamentally
    non-existing -
    $SU(5)$ symmetry.
    
    \subsection{The Plaquette Trace Action}
    As is usual, once you formulate your gauge theory on a lattice, you
    for smoothness reasons let the plaquette action typically be a linear
    function in the trace of the matrix representing the plaquette group
    element. This mainly from smoothness decided action will for the use
    of the reprenstation of the standard model group SMG
    on the ${\bf 5}$-plet function as if it were in $SU(5)$-thory.
    Actually it leads to couplings for the three sub-lie-algebras
    corresponding to the three Lie algebras $U(1)$, $SU(2)$ , and $SU(3)$
    being equal to each other in the same notation, in which they are
    equal in true $SU(5)$.
    So at first we have just from these simplicity and Taylor expansion type
    arguments gotten {\bf effective $SU(5)$ symmetry}!
    The plaquette action, as we shall use it to give the more precise
    reslult including also quantum fluctuations $H$, takes the form
    \begin{eqnarray}
      W_{\Box} &=& Re Tr \left ( \exp(i(h+H)) \right ),\label{tra} 
      \end{eqnarray}
    where both $h$ and $H$ are the Lie algebra valued fields written
    as represented by the representation on the ${\bf 5}$-plet.
    The $h$ symbolize the field for which we want to estimate an effective
    action, we can think of it as representing a continuous field translated
    into the lattice and matrx formulation. On the other hand the part $H$
    should
    describe the quantum fluctuations, i.e. quantum mechanically of course
    even in a situation, in which you classically describe the situation
    by the field from which $h$ has come. There is in reallity a superposition
    of fields configurations. That is to say, that the plaquette or link
    at a certain position in space time deviate appreciably from the
    confuguration given by $h$ which is the ``naive'' translation of the
    ansatz field considered to the lattice. It is this deviation we call
    $H$. In first approximation - and we shall be satisfied with that - the
    fluctuation part $H$ will be simply the fluctuation in vacuum.

    Now it is our calculational approach to Taylor expand the trace-action
    (\ref{tra}) to include the first term, which even on the average
    get non-trivial contribution from the fluctuations. We shall namely in
    our calculation show, that it is this lowest non-trivial order term in the
    fluctuations, which gives the {\bf deviation from $SU(5)$-symmetry.}
    And what really shall come out is, that indeed {\bf this contribtuion
      also fits
    with the deviation from $SU(5)$-symmetry} of the (inverse) fine structure
    constants as measured under use of the Standard Model.

    It is important for the present work to calculate, that the fluctuation
    in one component of $H$ is
    \begin{eqnarray}
      \frac{1}{2} <H_{one \; component}^2> &=&
      \frac{\pi}{2}\alpha\hbox{(in one layer lattice)}
      \end{eqnarray}
    and we shall do it in the following subsection. The reason we gave the
    value for 1/2 of the fluctuation, is that there is a factor 1/2 extra from
    the Taylor expansion, so that the counting of fluctuation contributuion
in the expression $Tr(H^2h^2)$ 
is to be multiplied by 1/2 to give the correction to the relevant
inverse fine structire constant. To the approximation that we in zeroth
orde have exact $SU(5)$ we do not have to distinguish, which precise one of the
various fine structure constants we shall use. This is also something, which
would require a bit more thinking / calculation and we would like to postpone
it for later article(see section \ref{ps11} ).

    Crucial is our Taylor expansion of the plaquette action (\ref{tra}),
    \begin{eqnarray}
      W_{\Box} &=& Re( Tr(\exp(i(h+H)) )\\
      &=& Re Tr[{\bf 1}] + Re (i Tr[ (h+H)] ) +\frac{1}{2}Re(- Tr[(h+H)^2])
      +...\nonumber\\
      && +\frac{1}{6}Re ( -i Tr [(h+H)^3]) + \frac{1}{24}Re Tr[(h+H)^4]
      +...\nonumber
      \end{eqnarray}
    The fields both the fluctuation $H$ and the ``test'' part $h$
    correspond
    to unitary representation matrices, are Hermitean as $5\times 5$
    matrices. Thus taking the real part removes the odd power terms, so they
    do not
    contribute, leaving
    in the above expansion up to 4th power,of interest, only the terms with
    2nd and fourth
    power. Now if we are interested in the corrections to the effective
    (continuum) finestructure constants, we only have interest in the terms of
    even order in $h$, and thus even from the fourth ordert term we only care
    for those six terms in the expnansion of $(h+H)^4$, which have two $h$
    factors and two $H$ factors. Among the a priori $2^4$ =16 terms in the
    $(h+H)^4$ development, there are only 6 terms with $h$ to just second
    power, and if the $h$ and $H$ commuted, these 6 terms would be identical.
    Indeed $h$ and $H$ do not commute, but when we take the average over the
    distribution
    of the fluctuations of $H$, it turns out that these terms after all
    have the same average, {\bf  as if} $h$ and $H$ did {\bf commute}.

    The terms to be kept for effective fine structure constant calculations
    purposes are:
    \begin{eqnarray}
      W_{\Box} &=& ...-\frac{1}{2}Re Tr[h^2] +...
      +\frac{1}{24}Re Tr[HHhh \hbox{(in any of 6 orders)}] +...
      \label{plae}\nonumber\\
      \hbox{if commuting}      &=& \frac{1}{2}( Re Tr [h^2] +
      Re \frac{1}{2}Tr[hhHH]+. 
    \end{eqnarray}
    In the last line we cancelled the factor 6 in the 24 by having here only
    one
    term, so this is achievable only, if the $h$ and $H$ ``effectively'' - i.e
    after averaging over the fluctuating $H$ - do commute.

    The full plaquette action shall have a coefficient $\beta$ in front of it,
    of course.
    To connect the continuum theory with
    Lagrangian density
    \begin{eqnarray}
      \hbox{\cal L}(x)&=& -\frac{1}{4g^2} F_{\mu\nu}(x)F^{\mu\nu}(x)\label{cl}a\\
      &=& -\alpha^{-1}/(16\pi)* F_{\mu\nu}(x)F^{\mu\nu}(x)
    \end{eqnarray}
    we should, say, using a normalization by
    \begin{eqnarray}
      F_{\mu\nu} &=& \partial_{\mu}A_{\nu}-\partial_{\nu}A_{\mu} + [A_{\mu},A_{\nu}]
    \end{eqnarray}
    identify for a link $h_-$ in the $\mu$ direction
    \begin{eqnarray}
      h_- &=& a^2A_{\mu}\\
      \hbox{and } h_{\Box} &=&\Sigma_{\hbox{around the box}}h_-
      \hbox{ (to linear approximation)} 
      \end{eqnarray}
    Calculationally it may be most easy to avoid problems with
    normalization to extract the {\bf ratio} of the second of these two terms to
    the first. The first of the two represent the naive(=lowest order)
    extraction of the continuum coupling from the lattice, while the second
    represents the lowest order effect of the fluctuations.
    
    \section{Extraction of Coupling Corrections}
\label{ps3}
    Once we have decided to look for the ratio of the second order and the
    fourth  order terms in the Taylor expansion of the Plaquette action 
    (\ref{plae}), we should be able to extract the {\em relative} correction
    due to inclusion of the quantum fluctuations by just putting in some ansatz
    for fields alone meaning a set up of one of the three standard model
    sub-group fields  at a time, and even the normalization
    (of $h$) is then not
    important for this
    relative size of the two terms, while the size of the fluctuations have
    to be calculated, though.

    Now we want to estimate the three Standard model finestructure constants
- or rather their ratios - by putting on a ``test field'' which for the
plaquette action, on which we think, is denoted $h=h_{\Box}$, and if we
think of
a purely spatial plaquette, is really a magnetic field of that plaquette.
This magnetic field is thought upon in the notation with the coupling
constant absored into the field, so that the action actually has an inverse
finestructure constant contained as factor to compensate the absorbed
charge-factor $e_0$ say,
\begin{eqnarray}
  S &=& ...+ \sum_{plaquettes} \frac{1}{2\pi\alpha_0} *Re Tr(U(\Box))\\
  \hbox{or continuum } S &\propto& \int \frac{1}{16\pi\alpha_0}
  F_{\mu\nu}F^{\mu\nu}d^4x.
\end{eqnarray}
(see section \ref{Wil} for why just $\frac{1}{2\pi\alpha}$ in front of
the $Re Tr(U_{\Box})$.)

Thus the inverse fine structure constant are found from how the action
(or say the magnetic energy) varies approximately linearly with the
square of the test field imposed $h^2$. If the fluctuation field was
$SU(5)$-invariant - as it would of course be in a theory without any
breaking of the SU(5)-symmetry, the three fine structure constants in the
``SU(5)'' invariant notation, which is wellknown to deviate from the
more natural one by the replacement:
\begin{eqnarray}
  \left .\frac{1}{\alpha_1}\right |_{natural} &=&
  \left .\frac{1}{\alpha_1}\right |_{SU(5)}*\frac{5}{3},
  \end{eqnarray}
would be equal to each other all three.

The test-fields, we shall use, and which for the non-abelian groups
$SU(2)$ and $SU(3)$ corresponds to the
coupling definitions
\begin{eqnarray}
  S &=&\int ( -\frac{1}{4e_2^2}\frac{1}{2}
  Tr_{matrix, 2\times 2}F_{\mu\nu}F^{\mu\nu })
  -\frac{1}{4e_3^2}\frac{1}{2} Tr_{matrix, 3\times 3}(F_{\mu\nu}F^{\mu\nu })+ ...)d^4x
  \nonumber
\end{eqnarray}
could be
\begin{eqnarray}
  \hbox{For $SU(2)$ : } h_{SU(2)} &=&\frac{1}{\sqrt{2}}\left [\begin{array}
      {ccccc}
      0&1&0&0&0\\
      1&0&0&0&0\\
      0&0&0&0&0\\
      0&0&0&0&0\\
      0&0&0&0&0
      \end{array}
       \right ],\\
  \hbox{for $SU(3)$ : } h_{SU(3)}&=&\frac{1}{\sqrt{2}}\left [\begin{array}
      {ccccc}
      0&0&0&0&0\\
      0&0&0&0&0\\
      0&0&0&1&0\\
      0&0&1&0&0\\
      0&0&0&0&0
      \end{array}
    \right ],\\
  \hbox{for U(1) : } h_{U(1)} &=& \frac{1}{\sqrt{30}}\left [ \begin{array}{ccccc}
      3&0&0&0&0\\
      0&3&0&0&0\\
      0&0&-2&0&0\\
      0&0&0&-2&0\\
      0&0&0&=&-2
      \end{array} \right ]\label{U1matrix}
    \end{eqnarray}

All the three proposed test-matrices $h$ have been normalized, so that their
squares
 \begin{eqnarray}
     h_{U(1)}^2 &=& \frac{1}{30}*diag(9,9,4,4,4)\\
     h_{SU(2)}^2&=& \frac{1}{2}diag(1,1,0.0.0)\\
     h_{SU(3)}^2 &=& \frac{1}{2}diag(0.0.,1,1,0)
   \end{eqnarray}
 become of trace equal to unity
 \begin{eqnarray}
   Tr(h_{U(1)}^2 ) &=& 1\\
   Tr(h_{SU(2)}^2 ) &=& 1\\
    Tr(h_{SU(3)}^2 ) &=& 1.
 \end{eqnarray}
 It is this normalization that ensures that the three couplings all become
 equal in the exact $SU(5)$ limit.
 (From the unbroken symmetry under the Standard model group it will not matter
 which component under one of the three standard model groups is used as test-
 field, as long as it is a combination of the components of just that one of the
 three groups $U(1)$, $SU(2)$ and $SU(3)$.)
 These fields $h$ are meant to be added to the already fluctuating
 field, but not to flutuate themselves, and then dividing the thereby
 achieved (magnetic) energy increase or action decrease we shall obtain
 (apart from a constant factor) the inverse finestructure constant for the
 subgroup of the Standard Model in question.

 \subsection{Difference Between Our Approximate $SU(5)$ and Usual $SU(5)$.}
\label{dbausu5}
 In the very first approximation - the $SU(5)$-invariant one - there is the same
 amount of
 fluctuation in all the 24 compponents of the $SU(5)$-Lie algebra,
 actually each of them have the average of the field squared for one
 component
 $1/2*<H_{\hbox{one component}}^2> = \frac{\pi}{2} *\alpha_5$.
 {\bf But in the
   philosophy, that only the Standard model components really exist, we
   must in our model only have fluctuations in these components.}

 The difference between our model, in which there truly speaking only is
 gauge symmetry by the Standard Model, and not even fields corresponding
 to the full $SU(5)$, and the usual $SU(5)$ theory comes in by {\bf restricting
   the fluctuation field $H$ in our model to {\bf only fluctuate in 
    Standard Model degrees of freedom.}}

 Actually the Lie algebra components, which are in the $SU(5)$-Lie-algebra
 but not in the Standard model one, can be in the notation, we have chosen here
 (\ref{U1matrix}), be represented by the matrix element being put to zero in the
 following matrix $5\times 5$:
 
 $\left [\begin{array}{ccccc}\cdot&\cdot&0&0&0\\
     \cdot&\cdot&0&0&0\\
     0&0&\cdot&\cdot&\cdot\\
     0&0&\cdot&\cdot&\cdot\\
     0&0&\cdot&\cdot&\cdot
   \end{array}\right ]$

 I.e. the difference between our model and the $SU(5)$ symmetric model is,
 that the fluctuation in the vacuum fields on the 2 times 6 points
 in this matrix marked by the $0$ 's is suppressed in our model, while
 in the $SU(5)$ symmetric $H$ the fluctuation is the same size in
 all the matrix element except for the detail that the trace of $H$ is
 restricted to be zero,
 \begin{eqnarray}
   tr(H) &=&0.
     \end{eqnarray}

 In both usual $SU(5)$ and ours  the trace is zero,but  the 12 element marked
 with zero
 are restricted from fluctuating only in our model.

 The technique to estimate what happens when one puts up in a
 region a smooth continium field is simply, that we add the field due to
 the continuum field, F say, translated to the matrix $h$ to the fluctuating
 field $H$. That is to say we consider the configuration:
 \begin{eqnarray}
   U(\Box) &=& \exp(i(H+h)),
   \end{eqnarray}
 then to extract magnetic energy or the action of the plaquette, we assume
 the usual type of real part of the trace action:
 \begin{eqnarray}
   S_{plaquette} &\propto& Re Tr(U(\Box)),
 \end{eqnarray}
 and look for the terms in the action change, which is 
 of  second order in the continuum extra field representing the continuum field.
 The coefficient to this second order $h^2$ to give the
 action change due to the continuum field is simply proportional to the
 inverse fine structure constant for the type of field we used.

 
\subsection{Expansion of $\exp(i(H+h))$}
 The Taylor expansion of the exponential is wellknown and we only have
 to keep the terms of second order in $h$, and we shall not go
 further than to second order in $H$, so we only need to expand to fourth
 order in the sum $H+h$. 

 
 In fact we generelly have
 \begin{eqnarray}
   Re Tr(\exp(i(H+h))&=& Re Tr({\bf 1})+ \frac{1}{2}Re Tr((i(H+h))^2)
   +\frac{1}{24}Re Tr((i(H+h))^4),\nonumber\\
   &&\hbox{(odd powers give zero)}.\nonumber\\
    \end{eqnarray}
 Dropping but the $h^2$ order terms we get
 \begin{eqnarray}
   \left.   S_{plaquette}\right |_{h^2-part} &=&\left.  Re Tr (U(\Box))
   \right |_{h^2-part}\\
   &=& \frac{1}{2}Re Tr(h^2)+\frac{1}{24}*6 Re Tr(h^2H^2)\label{twoterms}\\
   && \hbox{( provided that $h$ and $H$ commute)}\nonumber\\
   \hbox{Otherwise :}&=&\frac{1}{2}Re Tr(h^2)+\frac{1}{24}*
   \left (4 Re Tr(h^2H^2)+
   2Re Tr(hHhH)\right )\nonumber\\
   &=& \frac{1}{2}Re Tr(h^2)+\frac{1}{6}Re Tr(h^2H^2) + \frac{1}{12}Re Tr(hHhH).
 \end{eqnarray}
 
 \subsection{Classification of Fluctuations}
 For the presentation of the calculation of the quantum fluctuation
 corrections to the three different fine stracture constants in the Standard
 Model, we divide the fluctuations into four classes. Have in mind that in
 crudest approximation the vacuum fluctuations in the $SU(5)$ symmetric
 approximation consists of independent fluctuations after all the 24 basis
 vectors in a basis for the $SU(5)$ Lie algebra. Imaginig having chosen
 this basis so that the 12 basis vectors are also basis vectors for the
 three sub Lie algebras corresponding to the three Standard 
 Model groups, we can divide the fluctuation into four sets, denoted
 symbolically by $H_1$ for the fluctuation in the single mode of the
 $U(1)$ subgroup, $H_2$ for the fluctuation in the $SU(2)$ degrees of
 freedom, and $H_3$ for the $SU(3)$ fluctuations, and then for us the
 most interesting class $H_{int}$, namely those remaining fluctuations in
 the $SU(5)$ Lie algebra, which do not fall into any of the three welknown
 subgroups of $SU(5)$ in the Standard model, and which in our model are
 declared not to exist in Nature and thus must be removed. I.e. these
 fluctuations under the name $H_{int}$ are put to zero. With such a
 classification we can divide the fourth order term into a series
 in principle of $3\times 4$ combinations. In fact we can ask for any
 of the three finestructure constants for which we want to calculate the
 quantum fluctuation corrections, what the contribution is from one of any
 of the four fluctuation classses, $H_1$, $H_2$, $H_3$, and $H_{int}$.  

 \subsection{Calculation Description}

 We want to calculate the shift in the three inverse fine structure constants
 of the Standrd Model by first calculate the relative changes
 $\frac{\Delta \alpha^{-1}_i}{\alpha^{-1}_i}$ of these inverse
 finestructure constants $1/\alpha_i$ for $i = 1,2,3$ denoting respctively
 the subgroups $U(1)$, $SU(2)$, and $SU(3)$. Since we are now computing the
 ``correction'' after the very lowest order approximation is considered
 to be exact $SU(5)$ symmetry, we can in principle be careless with which
 finstructure constants we use in this calculation, when performed at the
 unification point of energy scale, because at this scale at zeroth
 approximation all three and even the $\alpha_5$ are equal.

 We shall first caculate the shifts $\Delta \alpha_i^{-1}(\mu_U)$ from their
 relative shifts. For this we need the very important
 $1/2*<H_{\hbox{one component}}^2>$
   = $\frac{\pi}{2}\alpha_5$ (but it is here we can be careless to our
   approximation with which $\alpha_1$ you replace this $\alpha_5(\mu_u)$),
   and the factor $\frac{\pi}{2}$ is explained below in section \ref{Wil}.

   Thus the shift of the inverse fine structure constant becomes
   \begin{eqnarray}
     \Delta \frac{1}{\alpha_i(\mu_U)} &=& \frac{1}{\alpha_i(\mu_U)} *
       \frac{ReTr(H^2h_i^2)}{2 Re Tr(h_i^2)}
       \hbox{(for effective commutativity)}\\
       &=&  \frac{1}{\alpha_i(\mu_U)} *<H^2_{\hbox{one component}}>*
       \frac{ReTr(H^2h_i^2)}{2 Re Tr(h_i^2)*<H^2_{\hbox{one component}}>}\nonumber\\
       &=&\frac{\pi}{2}*
       \frac{ReTr(H^2h_i^2)}{2 Re Tr(h_i^2)*<H^2_{\hbox{one component}}>}.
     \end{eqnarray}
   One can think of the fraction  $\frac{ReTr(H^2h_i^2)}{2 Re Tr(h_i^2)
     *<H^2_{\hbox{one component}}>}$ as a kind of counting how many components
   of the fluctuation contribute to the correction of the $i$th inverse fine
   structure constant,
   \begin{eqnarray}
     \hbox{``Eff. \# $<H^2>$ contributions''}&\stackrel{=}{def}&
     \frac{ReTr(H^2h_i^2)}{2 Re Tr(h_i^2)*<H^2_{\hbox{one component}}>}\\
     &=&\sum_{j=1,2,3}  \hbox{``Eff. \# $<H^2>$ contributions''}|_{H_j}.\nonumber
   \end{eqnarray}
   
   Here of course
   \begin{eqnarray}
     \hbox{``Eff. \# $<H^2>$ contributions''}|_{H_j}&\stackrel{=}{def}&
     < \frac{ReTr(H^2_jh_i^2)}{2 Re Tr(h_i^2)*<H^2_{\hbox{one component}}>}>\nonumber
     \end{eqnarray}
   If we include into this sum also the $H_{int}$ fluctuations, we get the
   corrections under unbroken $SU(5)$ and in this case the sum
   of these $\hbox{`` Eff. \# $<H^2>$ contributions''}$ should for
   all three inverse fine structure constants be $24/5$. There are
   24 components for full $SU(5)$, but in order to contribute to the trace
   $Tr$ a factor $1$ you need 5 $1$'s (along the diagonal).

   \subsection{The Table}
   By a little thinking of, that we want the average of these fluctuations which
   are independent, except along the diagonal, and that elements in the matrix
   related by permuting column number with row number are strongly correlated
   as must be the case to ensure hermiticity of the fluctuating fields
   $H=H^{\dagger}$, we find out that one gets the same result whatever the order
   in the matrix product, so that effectively $h$ and $H$ commute after all.

   Let us now list a table these $\hbox{```Eff. \# $<H^2>$ contributions''}$
   and their calculations:

   \begin{table}[h] 
     \caption{Table of the numbers $\frac{Re Tr(H_i^2h_j^2)}{2* Re Tr(h_j^2)
       <H_i^2>}$ first without the explicit denominator 2, but then at the
     very two lowest lines the half is taken for sum of the contribution
     from the Standard Model group fluctuations and for the ones from the
     $H_{int}$ which is missing in the standard model. \label{tabcorr}}
     \begin{center}
       
   \begin{tabular}{|c|c|c|c|}
     \hline
     From&$\alpha_1^{-1}$& $\alpha_2^{-1}$&$\alpha_3^{-1}$\\
     the $H_i$&$h_{U(1)}$&$h_{su(2)}$&$h_{SU(3)}$\\
     \hline
     \hline
     $H_1$&$ \frac{2*81+3*16}{900}$&$\frac{2*9}{2*30}$&$\frac{4}{30}$\\
     &=7/30&=3/10&=2/15\\
     \hline
     $H_2$&$\frac{3*9*2}{3*30}$&$\frac{2*3}{2*2}$&0\\
     &=9/10&=3/2&=0\\
     \hline
     $H_3$&$\frac{4*3*8}{3*30}$&$0$&$\frac{8}{3}$\\
     &=16/15&=0&=8/3\\
     \hline
     \hline
     sum& 11/5& 9/5&14/5\\
     \hline
     \hline
    $ H_{int}$& $\frac{54+24}{30}$&$3$&$2$\\
     &=13/5&=3&=2\\
     \hline
     \hline
     check & 24/5&24/5&24/5\\
     \hline
     \hline
     half s.&11/10&9/10&7/5\\
     half $H_{int}$&13/10&3/2&1\\
     \hline
     \hline
   \end{tabular}
   
   \end{center}

   \end{table}

 The numbers in this table are easily obtained when having in mind when the
 trace is of the form $Tr(H^2h^2)$ because we can then simply evaluate the
 traces by using the following diagonal matrices:
 \begin{eqnarray}
   <H_1^2> &=& \frac{1}{30}*diag(9,9,4,4,4)\\
   <Tr(H_1^2)> &=& 1\\
   <H_2^2> &=& \frac{3}{2}*diag(1,1,0,0,0)\\
     <Tr(H_2^2)> &=& 3\\
     <H_3^2> &=& \frac{8}{3}diag(0,0,1,1,1)\\
     <Tr(H_3^2)> &=& 8\\
     <H_{int}^2> &=& diag(3,3,2,2,2)\\
     <Tr(H_{int}^2)> &=& 12
     \end{eqnarray}
 combined with the squares of the ansatz matrices
 \begin{eqnarray}
   h_{U(1)}^2 &=& \frac{1}{30}diag(9,9,4,4,4)\\
   Tr(h_{u(1)}^2) &=& 1\\
   h_{SU(2)}^2 &=& diag(1/2,1/2,0,0,0)\\
   tr(h_{SU(2)}^2) &=& 1\\
   h_{SU(3)}^2 &=& diag(0,0,1/2,1/2,0)\\
   Tr(h_{SU(3)}^2)&=& 1
 \end{eqnarray}
 
\subsection{The Problem with Commutation}

The above multiplication to make the table is o.k. if the $h$'s and $H$'s
indeed commute. Effectively, however, we can show that by the averaging,
we do end up as if they commuted:

The $h$'s, i.e. the ansatz matrices, we can simply choose diagonal, because
that is just to select an appropriate basis vector for the group one wants.
If the fluctuation field is a diagonal one it is then indeed commuting, but
if we consider an off-diagonal component of an $H_i$ field, then we can argue
that it leads to a product of the two diagonal elements in the $h$
and this leads in the special cases we consider to taking trace of an $h$
which is zero. So in pracsis it is as if we had commutation, almost by accident.

 \section{Wilson Action}
    \label{Wil}
    We shall use the notation for the single layer (our model
    has three layers corresponding to three families) Wilson lattice, being
    related to a continuum
    theory(we here leave the gauge group open) and with the charge
    absorbed into the field $F^{\mu\nu}(x)$ (containing magnetic $\vec{B}$
    and electric part $\vec{E}$ with their $g$ absorbed):

If we use a notation, in which the $A_{\mu}(x)$ gauge fields are
already Lie-algabra valued fields - or for our $U(N)$ groups of interest here
equivalently matrices - and thus can define basis-vector matrices $\lambda_a$
and $T_a$ so that 
\begin{eqnarray}
  A_{\mu}(x) &=& (\Sigma )  A_{\mu}^a\frac{\lambda_a}{2} \\
  &=& (\Sigma ) A_{\mu}^a T_a\\
  \hbox{where, say, for off-diagonal } \lambda_1 &=&
  \left [\begin{array}{ccccc}0&1&0&0&0\\
     1&0&0&0&0\\
     0&0&0&0&0\\
     0&0&0&0&0\\
     0&0&0&0&0
    \end{array}\right ]\\
  \lambda_2&=&  \left [\begin{array}{ccccc}0&-i&0&0&0\\
     i&0&0&0&0\\
     0&0&0&0&0\\
     0&0&0&0&0\\
     0&0&0&0&0
    \end{array}\right ]\\
  \hbox{and with normalization } Tr(\lambda_a \lambda_b) &=& 2 \delta_{ab}\\
  \hbox{and } Tr(T_aT_b) &=& 1/2 *\delta_{ab}
  \end{eqnarray}
you can by interpreting the $A_{\mu}(x)$ fields as representation in some
representation $R$ construct unitary matrices in the crude continuum limit
identification
\begin{eqnarray}
  U_{\mu}(x) &=& \exp(ia A_{\mu}(x))
\end{eqnarray}
in the usual way require the
\begin{eqnarray}
  S_{Wilson}[U] &=&- \frac{\beta}{2N}\Sigma_{\Box}(W_{\Box}+W_{\Box}^*)\label{Wn}\\
  &=& \frac{a^4\beta}{4N}\int \frac{d^4x}{a^4} trF_{\mu\nu}F_{\nu\mu} +...\\
  \hbox{where }  
  W_{\Box}&=& tr(U_{\mu}(x)U_{\nu}(x+\hat{\mu})U^{\dagger}(x+\hat{\nu})
  U^{\dagger}(x))\\
  &=& tr(\hbox{oredered product around the plaquette $\Box$})\nonumber
\end{eqnarray}
obtain using (\ref{cl})
$S = \int -\frac{1}{4 g^2} F_{\mu\nu}F^{\nu\mu}d^4x $ the relation
\begin{eqnarray}
  \frac{\beta}{2N}&=& \frac{1}{g^2}.\label{b2}\\
  \hbox{or } \frac{\beta}{N} &=& \frac{1}{2\pi\alpha}\label{b2n}.
\end{eqnarray}

And this leads to that the fluctuating part $H=(\Sigma) H^aT_a =(\Sigma)
H^a\frac{\lambda_a}{2}$ of the exponent in the plaquette variable
\begin{eqnarray}
  U_{\Box} &=& \exp(i \Sigma H^a\frac{\lambda_a}{2})
  \end{eqnarray}
 goes into the action with
 \begin{eqnarray}
   \Sigma_{\Box} \frac{\beta}{N} Re tr \exp(i \Sigma H^a\frac{\lambda_a}{2} )\\
   &=& \Sigma_{\Box} \frac{1}{2\pi\alpha}Re tr  \exp(i
   \Sigma H^a\frac{\lambda_a}{2})\\
   &\stackrel{\approx}{\hbox{second o.}}&  \frac{1}{2\pi\alpha}\Sigma_{\Box}
   Re tr (-\frac{1}{2}(\Sigma_aH^a\frac{\lambda_a}{2})^2)\\
   &=&\frac{1}{4\pi\alpha}\Sigma_{\Box} \Sigma_a (H^a)^2 /2\\
   &=& \Sigma_{\Box \; a} \frac{1}{8\pi \alpha} (H^a)^2
   \end{eqnarray}
 So if the plaquettes were not coupled - what they though are - then
 in the partition function /
the Euclidean path integral which is
\begin{eqnarray}
Z &=& 
\int
DU \exp(-\beta S[U])\\
&\approx & \Pi_{\Box \; a} \exp(-\frac{1}{8\pi\alpha}*( H^a)^2)
\end{eqnarray}
where $DU$ is the Haar measure,
 the fluctuation of a plaquette variable (exponent) $H^a$ would be given as
 $<(H^a)^2> \hbox{(no summation)}$ =$ 8\pi\alpha/2$ (when restriction between
 the plaquette variables were neglected),
 since $\frac{\int x^2 \exp(-Kx^2)dx}{\int \exp(-Kx^2) dx} = 1/(2K)$. But of
 course they are connected
 so that there are only half the plaquette variables,which are independent.
 This can actually be seen to lead to that the distribution of the partition
 function distribtuion become twise as narrow measure in the square $H^a$
 average: So in the lattice partition faction or the Euklideanized path
 integral the fluctuation is
 \begin{eqnarray}
   <(H^a)^2> \hbox{(no summation)} &=& 8\pi\alpha/2/2 = 2\pi \alpha.
   \end{eqnarray}

We here used that the plaquette variables, say $H^a( \Box)$ for the different
plaquettes $\Box$ are not independently integrated over. On the contrary
for each cube in the lattice there is a constraint which linearized means
that the sum of six plaquaette variable for the plaquettes around the
cube is restricted to be zero. Since there in 4 dimensions are 6 plaquettes per
site and 4 cubes, this restriction would in first go mean that there per
site were only 2 independent plaquette variables, but that is, however,
not true, because there is a constraint between the four cube-constraints on
the plaquettes. So in reality there is per site 3 independent constrains on 6
a priori plaquette variables. This gives that the average of the square
$(H^a)^2$
of a (Gaussian distributed) plaquaette variable get reduced by a factor
6 to (6-3) meaning a factor 2. Simplifying to just 2 variables to get
restricted to 1 independent we could just think of a Gaussian distribution
about the origo in a plane, and that we then restrict the at first two
dimensions to a diagonal - a single dimension - being a restriction
symmetric between the two orinal variables thought of as the coordinates.
Then the restricted distribtuion on the symmetric diagonal would project into
one of the coordinate axes with the average of the saquare diminished by a
factor 2.

The meaning of our basis choice for defining our lattice variables $H^a$
could be illustrated by asking, what is now the calculated average of the
square of an off diagonal element in the $5 \times 5$ matrix. E.g. for
matrix element row 1 column 2 we get
\begin{eqnarray}
  <|H_{\hbox{row 1 column 2}}|^2> &=&< (H^1/2)^2 + (H^2/2)^2>\\
  &=& 1/2*2\pi \alpha = \pi \alpha.
\end{eqnarray}

It is such an - most easy off diagonal element we denote
by $H_{\hbox{one component}}$ and its numerical  average square is thus for
{\bf one layer}
\begin{eqnarray}
  <|H_{\hbox{one component}}|^2>|_{\hbox{one layer}}&=&\pi \alpha.\\
  \hbox{Want } \frac{1}{2} <|H_{\hbox{one component}}|^2>|_{\hbox{one layer}}&=&
  \pi/2* \alpha.
\end{eqnarray}
The reason we want this half of the average square of the matrix element
in the $5\times 5$ matrix, is that in the Taylor expansion (\ref{twoterms})
has a factor 2 deviation between the two terms,which we shall compare.

\subsection{Our Relative Correction}

In the calculation of the relative correction to the inverse
exact $SU(5)$ fine structure constants we need the ratio of the two terms
(\ref{twoterms}) and the correction term comes from the Taylor expansion
as
\begin{eqnarray}
  \hbox{`` correction term''} &=& \frac{1}{4}* tr( h^2H^2)\hbox{(if
    commuting effectively)}\nonumber\\
  \hbox{while the corresponding ``uncorrected''  } &=&\frac{1}{2} tr(h^2). 
 \end{eqnarray}
\subsection{On Table}
  Use the numbers from thee table being just traces of the products of the
  diagonal matrices, which are normalized so that their traces are $1$
  for the $h^2$ and the dimension of the Lie Group for the $H_i^2$ - normalizes
  the difference $\frac{1}{\alpha_2} -\frac{1}{\alpha_3}$ to one ``unit''
  ignoring yet the factor 3 of number of families, and the hereby absorbed
  denominator 2, being
  \begin{eqnarray}
    \hbox{ The ``unit'' } &=& \frac{\pi}{2}
    \end{eqnarray}
  now in the notation with ``Re Tr'' (in which it would at first have been $\pi
  $).
  So the prediction will be that the difference at the unifying scale of the
  two nonabelian inverse fine strucutre constants - which had number 1
  (when the explicit 1/2 not included) - will be $\frac{\pi}{2}$ for
  only one family, but
  $3*\pi/2$ for three families.

  \section{Compare with Old Work with Bennett, and with Computer Works}
  \label{ps6}
  Since it is so crucial for our prediction that we calculate the absolute
  size of the quantum correction,our $q=3*\pi/2$ correctly and that it is
  indeed such a quantum correction effect,we shall here compare it to an old
  work  with Don Bennett, though only calculating this correction for
  simple groups SU(3) and SU(2), but it checks the absolute size. That the
  physics of this type of quantum correction works even with a background
  of an extensive computer calculation is seen in the next subsection
  \ref{NY}
  In my old work with Don Bennett  
\cite{Don93} arXiv:hep-ph/9311321v1 
``Predictions for Nonabelian
Fine Structure Constants
from Multicriticality'' we in fact presented the same correction, which
we use here and even had the normalization included and used that the
correction to the inverse fine structure constants are
\begin{eqnarray}
  \frac{1}{\alpha} &\rightarrow& \frac{1}{\alpha}(1-C_f\pi\alpha)\\
  &=& \frac{1}{\alpha}-C_f\pi
\end{eqnarray}
where $C_f$ means the quadratic Cassimir in the fundamental representation
of the group in question. In fact we find in this article:
\begin{eqnarray}
  C_f^{SU(2)} &=& \frac{3}{4}\\
  C_f^{SU(3)}&=& \frac{4}{3}
\end{eqnarray}

\subsection{Tadpole Correction Calculations}
\label{NY}
In fact the quatity $<H_i^2>$, which is so crucial to us to  get estimated,
is a quantity needed to make the so called tad-pole improvements for lattice
calculations\cite{tadpole}. In the calculation by Niyazi eta. \cite{Niyazi}
we find some computer study, that also reach the quantity $u_0$ defined by
\begin{eqnarray}
  u_0^4 &=& \left < \frac{1}{N} Tr(U_p(\Box)) \right >, 
  \end{eqnarray}
or as being the average value in the fluctuating lattice (in vaccum) for a
link variable. They present as a result of their numerical studies in a region
of $\beta$'s around $\beta=7.5$ in their nottation meaning $1/\alpha_3 = 7.5/5
*2 \pi = 9.42477$:
\begin{eqnarray}
  u_0(\beta) &=& 0.87010 + 0.03721 \Delta \beta -0.01223 (\Delta \beta)^2.
\end{eqnarray}
where $\Delta \beta = \beta -7.3$.

On the basis of the  crudest approximations as speculated in our section
\ref{Wil} we expect the $u_0(\beta)$ to be of the form
  \begin{eqnarray}
    u_0^4(\beta)&=& 1 - \frac{C}{\beta}\\
      \hbox{needing then } C&=& 7.3*(1 - 0.87010^4)\\
      &=& 7.3*(1-0.057316)\\
      &=& 3.1159.\\
      \hbox{If so, shift } \Delta \frac{1}{\alpha_3}&=& C/3 *2\pi*
      (1- \frac{2*4}{20})\\
&=&      C*2\pi/5 \\
      &=& 3.9155\\
      &\approx & 4.1888\\
      &=& 8/3*\pi/2
  \end{eqnarray}
  (here the correction factor comes from our (\ref{b2}) , correction
  for a $N_c=3$ in the notation of Niyazi et a.,and correction because the
  continuum coupling - the $\alpha_3$ -gets a contribution from a lattice
  action term with double plaquettes having a coefficient $\beta/20$
  in first approximation and contribtuiing 8 times as much as the ``main
  Wilson term'')
  If the inverse $\beta$ type fitting here is correct, then the derivative
  being the coefficiient on the second term $0.03721\Delta \beta$ should be
  \begin{eqnarray}
    \frac{d}{d\beta}u_0(\beta)&=& \frac{d}{d\beta} \sqrt[4]{1-\frac{C}{\beta}}\\
      &=&\frac{1}{4} (1-\frac{C}{\beta})^{-3/4}*(\frac{C}{\beta^2})\\
      &=&\frac{1}{4}u_0^{-3}*C/\beta^2\\
      &=& \frac{1}{4}C/7.3^2/0.87010^3\\
      &=& C*0.00712\\
      &=& 0.022190.
  \end{eqnarray}
This is a little bit lower than the $0.03721$.

From formula (2) in reference \cite{Niyazi} we see that Niayzi et al.
uses the $N$ included in the action explicitely so that for $SU(3) $ their
$\beta = 3\beta_{without \; the N}$,so e.g.  the $\beta=7.3$ where
they worked would mean in the notation without the $N$ included in the
definition $7.3/3 =  2.4333$. Then since in the usual notation
which Niayzy et al.seems to use one has e.g. according to \cite{Vege}
$\beta =\frac{2N_c}{g_s^2}$ implying
\begin{eqnarray}
  \frac{1}{\alpha_3}&=& \frac{4\pi}{g_s^2}= 2\pi \beta_{\hbox{with no $N_c$ notation}}
\end{eqnarray}

But there is a further point in extracting the fine structure constant used in the
work by Nyaizi et al: They use L{\''u}scher-Weisz action which even in the
large $\beta =\beta_{pl}$ limit has an extra term consisting of double
plaquette actions with a coefficient which according to \cite{Alford}
is given by the
\begin{eqnarray}
  \beta_{rt}&=& -\frac{\beta_{pl}}{20 u_0^2}*(1+0.4905\alpha_3)\\
  S[U] &=& \beta_{pl} \Sigma_{rt}\frac{1}{3} Re Tr(1-U_{pl})\\
  &+& \beta_{rt}\Sigma_{rt}\frac{1}{3}Re(1-U_{rt})\\
  &+& \beta_{pg}\Sigma_{pg}\frac{1}{3} ReTr(1-U_{pg})\\
  \hbox{so } \beta_{eff}|_{lowest\; order} &=& \beta_{pl}*(1-\frac{1}{20}*4*2)\\
  &=& \beta_{pl}*\frac{3}{5}
  \end{eqnarray}
So this would mean we shall use  (\ref{b2n}), but with
$\beta/N$ put to $3/5*\beta_{pl}/3$. The case $\beta=7.3$ in the
notation of
\cite{Niyazi} corresponds then to
\begin{eqnarray}
  2\pi* 2.433333 &=& \frac{1}{\alpha_3}\\
  \hbox{giving }1/\alpha_3 &=&
  15.2890 \hbox{ forgetting the 3/5}\nonumber\\
  \hbox{so the } u_0^4 &=& 0.87010^4= 0.573161057\\
  \hbox{will correct  by}
  15.2890*(1-0.87010^4)
  &=&6.25597
  \\
  \hbox{which should be }\pi/2 *8/3
  &=& 4.18878
\end{eqnarray}

However,when we now remember the inclusion of the effect of the
double plaquette term at least in the weak coupling limit giving the
factor 3/5, then instead to Niyazi et al. 's $\beta =7.3$:
\begin{eqnarray}
  \beta_{true} &=& 7.3/3 *3/5\\
  &=& 7.3/5 \\
  &=& 1.46\\
  \hbox{giving }\frac{1}{\alpha_3} &=& 2\pi \beta_{true}\\
  &=& 9.1734\\
  \hbox{and shift by } 9.1734*(1-0.87010^4) &=& 3.91556\\
  \hbox{again compare to } 8/3*\pi/2 &=& 4.188787
  \end{eqnarray}

Now there is very little difference,so that we can consider it, that this
extraction from the calculation of the $u_0$ became a test of our
calculation of the correction from loop corrections being so crucial
for the present work.

Let us take yet an example namely $\beta_{Nyaizi} =7.7$,it gives
$\beta = \beta_{Nyaizi} /3*3/5 =
1.54$ and
$1/\alpha_3 = 2\pi *
1.54 =9.6761 $.
Now we had for 7.7, $u_0=0.8803$
and thus $1-u_0^4 =1 -0.8803^4=0.399486$ giving  
the change of the $9.6761$ by $3.8655$.
Still close to $4.1887$ (But I do not like it got further
away from this 4.1887, when the coupling got weaker, because
we expect our values exact in the weak coupling limit).


\section{Fitting}
\label{ps7}
The first step in our fitting of our model  is to calculate the ``unifying''
scale $\mu_u$, at which the ratios between the differences between the inverse
fine structure constants for the three subgroups of the Standard Model group
is the one predicted from our calculation of the quantum fluctuation
corrections. In fact the three inverse fine structure constants shall lie on
the number axis as the numbers $(2,13/5,3)$ corresponding to the
subgroups $(SU(3), U(1), SU(2))$, where we have chosen the
$SU(5)$-normalisation for the $U(1)$-finstructure constant. The relation is
expressed in terms of the two independent differences, that can be formed.
Let us , e.g., say
\begin{eqnarray}
  \frac{\frac{1}{\alpha_2} - \frac{1}{\alpha_{1\; SU(5)}}}{3-13/5}&=&
    \frac{\frac{1}{\alpha_{1 \; SU(5)}} - \frac{1}{\alpha_3}}{13/5 -2}\\
        \Rightarrow  \frac{1}{\alpha_2} - \frac{1}{\alpha_{1\; SU(5)}}&=&
        \frac{2}{3}*(\frac{1}{\alpha_{1 \; SU(5)}} - \frac{1}{\alpha_3})\\
        \Rightarrow \frac{1}{\alpha_2}-\frac{5}{3}*\frac{1}{\alpha_{1\; SU(5)}}+
          2/3*\frac{1}{\alpha_3} &=&0
\end{eqnarray}
Expressing the $\frac{1}{\alpha_i}$'s as
\begin{eqnarray}
  \frac{1}{\alpha_i(\mu)} &=& \frac{1}{\alpha_i(M_Z)} -\frac{b_i}{2\pi}
  \ln\left (\frac{\mu}{M_Z} \right )+...\\
  \hbox{with } 
b_i^{SM}& =& (41/10, -19/6, -7), 
\end{eqnarray}
this relation for the $\alpha_i(\mu_u)$ 's is written for the $M_Z$-scale
finestructure constants as
\begin{eqnarray}
  \frac{1}{\alpha_2(M_Z)}-\frac{5}{3}*\frac{1}{\alpha_{1\; SU(5)}(M_Z)}+
  2/3*\frac{1}{\alpha_3(M_Z)} &=&(b_2-\frac{5}{3}b_1+2/3*b_3)/(2\pi)
  *\ln \left(\frac{\mu_u}{M_Z}  \right ).\nonumber
\end{eqnarray}

Inserting the values obtained for the $M_Z$ inverse fine structure
constants this becomes:
%
\begin{eqnarray}
  29.57\pm 0.06\% -\frac{5}{3}*59.00\pm 0.02\% +\frac{2}{3}* 8.446 \pm 0.6\%
  &=& \frac{-19/6 -5/3 *41/10 +2/3*(-7)}{2\pi}*\nonumber\\
  &*& \ln \frac{\mu_u}{M_Z}\nonumber\\
  -63.10&=& -44/3/6.2832\ln\frac{\mu_u}{M_Z}\\
  \Rightarrow \ln\frac{\mu_u}{M_Z} &=&
  27.03\\
  \Rightarrow \frac{\mu_u}{M_Z}&=&
  5.482 *10^{11}\\
  \hbox{Using } M_Z &=&91.1876 GeV\\
  \hbox{thus } \mu_u &=&
  5.00*10^{13}
  \end{eqnarray}
   
\subsection{Table for Inverse Fine Structure Constants and Our Fitting}

In the table 6.1
we go through the calculation of first determine
the our unification scale by requiring the ratios of the two relative
deviations from true $SU(5)$ symmetry to be in the ratio required by our
model. This we have shown to  be done by requiring the linear combination
of the three inverse finestructure constants at this unifying scale to
make zero the linear combination of the inverse fine structure constants
having the coefficients $(-5/3, 1, 2/3) $ for respectively $(1/\alpha_{1\; SU(5)
}, 1/\alpha_2, 1/\alpha_3)$. As a check of our model we work out, by
correcting for the quantum fluctuations in the inverse fine structure constant,
to reproduce the two $1/\alpha_5$'s, namely the one without quantum corrections
- the bare $SU(5)$ inverse fine structure constant - and the `effective ''
$SU(5)$ inverse fine structure constant which has been corrected for these
quantum corrections. The test is that these two formal$SU(5)$ (inverse)
couplings shall be the same whichever of the three standard model fine
structure constants are used for the calculation of them, provided
our model agrees with the data used.

      \begin{table}[!]
     
        \begin{adjustbox}{max width = \textwidth}{
         \label{ta}
      \begin{tabular}{|c|c|c|c|c|}
        \hline
        &$1/\alpha_{1 \; SM}$& $1/\alpha_{1 \; SU(5)}$&$1/\alpha_2$ & $1/\alpha_3$
        \\
        \hline
        Formula & $1/\alpha_{EM}cos^2\Theta_W$& $3/5*1/\alpha_{EM} cos^2\theta_W$&
       $ 1/\alpha_{EM}*sin^2\Theta_W$&$\alpha_3^{-1}$
        \\
        \hline
        Start $\#'s$&$ 127.916 *0.76884$&$\frac{3}{5}*127.916 *0.76884$ &
       $ 127.916*0.23116$&$0.1184^{-1}$\\
        \hline
        Value & 98.347 & 59.008 &29.569&8.446\\
        Uncertainty &$\pm$ 0.02& $\pm$ 0.013 & $\pm$ 0.017&$\pm$ 0.05\\
        \hline
        Coefficient&&-5/3&1&2/3\\
        \hline
        Contribution&& -98.347 & 29.569 &5.631\\
        Uncertainty&&$\pm$ 0.02& $\pm$ 0.017&$\pm$ 0.034\\
        \hline
        &{\bf SUM:}&&&\\
        Sum &-63.147&&&\\
        Uncertainty& $\pm$ 0.04&&&\\
        \hline
        b's&41/6&41/10&-19/6&-7\\
        \hline
        b-contribution& & -5/3*41/10 & 1*(-19/6)& 2/3*(-7)\\
        && = -41/6& = -19/6 & =-14/3\\
        \hline
        Sum &(-41-19-28)/6&&&\\
        &=-44/3&&&\\
        \hline
        b-contr./$2\pi$& -2.33420017&-1.087559696&-0.503991079&-0.742723695\\
        \hline
        &{\bf Ratio:}&&&\\
        $\ln(\frac{\mu_U}{M_Z})$&$\frac{-63.147}{-2.33420017}$&&&\\
        &={\bf 27.053}&&&\\
        Uncertainty&$\pm$ 0.02&&&\\
        \hline
        Scale $\mu_U$& $5.116*10^{13} GeV$&&&\\
        Uncertainty&$\pm 0.1*10^{13} GeV$&&&\\
        \hline
        \hline
        b's/$2\pi$&&0.652535818& -0.503991079&  -1.114085543\\
        \hline
        $\ln(\frac{\mu_U}{M_Z})*\frac{b's}{2\pi}$&&17.653
        & -13.634&-30.139\\
        Uncertainty&&$\pm$ 0.01&$\pm$ 0.01&$\pm$ 0.02\\
        \hline
        Value at $\mu_U$& & 41.355& 43.203& 38.585\\
        Uncertainty&&$\pm$ 0.017& $\pm$0.02&$\pm$ 0.05\\
        \hline
        Pred.to $1/\alpha_{5\; bare}$&&3*11/5*$\pi/2$&3*9/5*$\pi/2$
        &3*14/5*$\pi/2$\\
        &&=10.367247&=8.482293&=13.194678\\
        \hline
        $1/\alpha_{5\; bare}$&& 51.722322462 & 51.685853652 &  51.780040772\\
        Uncertainty&&$\pm$ 0.017&$\pm$ 0.02&$\pm$ 0.05\\
        \hline
        Pred. to$1/\alpha_{5\; cont}$&&3*13/5*$\pi/2$&3*3*$\pi/2$&3*2*$\pi/2$\\
        &&= 12.252201&=14.137155&=9.42477 \\
        \hline
        $1/\alpha_{5 \; cont}$&&29.103&29.066& 29.161\\
        Uncertainty&&$\pm$ 0.017&$\pm$ 0.02&$\pm$ 0.05\\
        \hline
        &{\bf Average:}&&&\\
        Average&29.092&w=35&w=25&w=4\\
        Deviations&&0.0107&-0.0258& 0.0683\\
        \hline
        \end{tabular}}
      \end{adjustbox}
\end{table}
      

        
        \subsection{Values at the $\mu_u$-Scale}
        What we are really interested in is the magnitude of the deviation
        from $SU(5)$ being accurate at the our ``unified scale'' $\mu_u$,
        and we should like to develop the expression for this deviation in
        terms of the original variables at $M_Z$ even. But to get an
        overview it is better first obtain the deviations by simply
        calculating the three inverse finstructure constants at the our
        ``unification scale'' $\mu_u$:
        \begin{eqnarray}
          \frac{1}{\alpha_{1\; SU(5)}(\mu_u)} &=&59.00\pm 0.02-0.65254*27.0566\\
            &=& 59.00 -17.66\\
          &=& 41.34\\
          \frac{1}{\alpha_2(\mu_u)} &=& 29.57 +0.50399*27.0566\\
          &=& 29.57+13.64\\
          43.21\\
          \frac{1}{\alpha_3(\mu_u)} &=& 8.446 +1.11409*27.0566\\
            &=& 8.446+30.143\\
            &=& 38.59
        \end{eqnarray}

        We may note down the differences and check that they are in
        the right ratio:
        \begin{eqnarray}
          \frac{1}{\alpha_2(\mu_u)} - \frac{1}{\alpha_{1\; SU(5)}(\mu_u)} &=&
          43.21 -41.34\\
          &=& 1.87.\label{d21}\\
          \frac{1}{\alpha_{1\; SU(5)}(\mu_u)} -\frac{1}{\alpha_3(\mu_u)}
            &=& 41.34 - 38.59\\
            &=& 2.75\label{d13}\\
            \frac{1}{\alpha_2(\mu_u)}-\frac{1}{\alpha_3(\mu_u)}
            &=& 43.21-38.59\\
            &=& 4.62\label{d23}
          \end{eqnarray}
        The test now is if
        \begin{eqnarray}
          2/5*4.62 &\stackrel{?}{=}& 1.87\\
          \hbox{In fact } 2/5*4.62&=&1.85\\
          \hbox{and }3/5*4.62& \stackrel{?}{=} &2.75\\
          \hbox{In fact } 3/5*4.62 &=& 2.77
        \end{eqnarray}

        Now our question is how big is this 4.62 in units of $\pi/2 =1.5708$.
        We find

        \begin{eqnarray}
          \frac{\frac{1}{\alpha_2(\mu_u)} -\frac{1}{\alpha_3(\mu_u)}}{\pi/2}\\
          &=& \frac{4.62}{\pi/2}\\
          &=& 2.94 \approx 3 = \# families !
          \end{eqnarray}
        This is remarkably close to 3, the number of families!
        (with an order of magnitude uncertainty $\pm 0.1$ it the inverse
        finestructure constants, a deviation of only $0.06$ is very good!)
        This is in itself a remarkable coincidence, in spirit with
        our old work stories about critical inverse finestructure constants
        getting multiplied by the number of families, because of the
        antiGUT theory behind.

        Corresponding to this spacing, we can now with the above
        calculations being used find in fact two $SU(5)$ inverse couplings,
    namely one before the effect of the quantum fluctuations $<H^2>$  of $H$ are
    taken into account and one after they are taken into account  for the
    - in our theory non-existent - whole $SU(5)$.


    \subsection{The $SU(5)$ Unification Couplings}
    
    Using the table 6.1
    we find, that using as unit
    $1/\alpha_2(\mu_u) -1/\alpha_3(\mu_u) = 4.62 \approx 3\pi/2$, the two
    a bit different inverse unified couplings $1/\alpha_{5\; bare}$ and
    $1/\alpha_{5\; cont}$ (for $SU(5)$ formally) at our unification scale
    $\mu_u$ are given
    as
    \begin{eqnarray}
      \hbox{The ``bare'':} &&\nonumber \\
      1/\alpha_{5\; bare} &= & 1/\alpha_{1\; SU(5)}(\mu_U) +11/5* 4.62= 41.34
      +10.164=51.504\nonumber\\
      & \hbox{or}& 1/\alpha_2(\mu_u) + 9/5*4.62=43.21 + 8.316 =51.526 \\
      & \hbox{or}& 1/\alpha_3(\mu_U) + 14/5*4.62=38.59+ 12.936 = 51.526\\
      \hbox{The corrected:}&&\nonumber \\
      1/\alpha_{5 \; cont}(\mu_U) &=& 1/\alpha_{1\; SU(5)}(\mu_U) - 13/5*4.62
      = 41.34 - 12.012 = 29.328\nonumber\\
      &\hbox{or }& 1/\alpha_2(\mu_U) - 3*4.62 = 43.21-13.86 = 29.35\\
      &\hbox{or }& 1/\alpha_3(\mu_U) - 2*4.62 = 38.59 -9.24=
      29.35
    \end{eqnarray}

    (We used in this table the ``experimental'' value $q=4.62$ but it
    would have made only very little difference to use the theoretical
    value $q=3*\pi/2$, because our agreement is so good)
    
\includegraphics[scale=0.8]{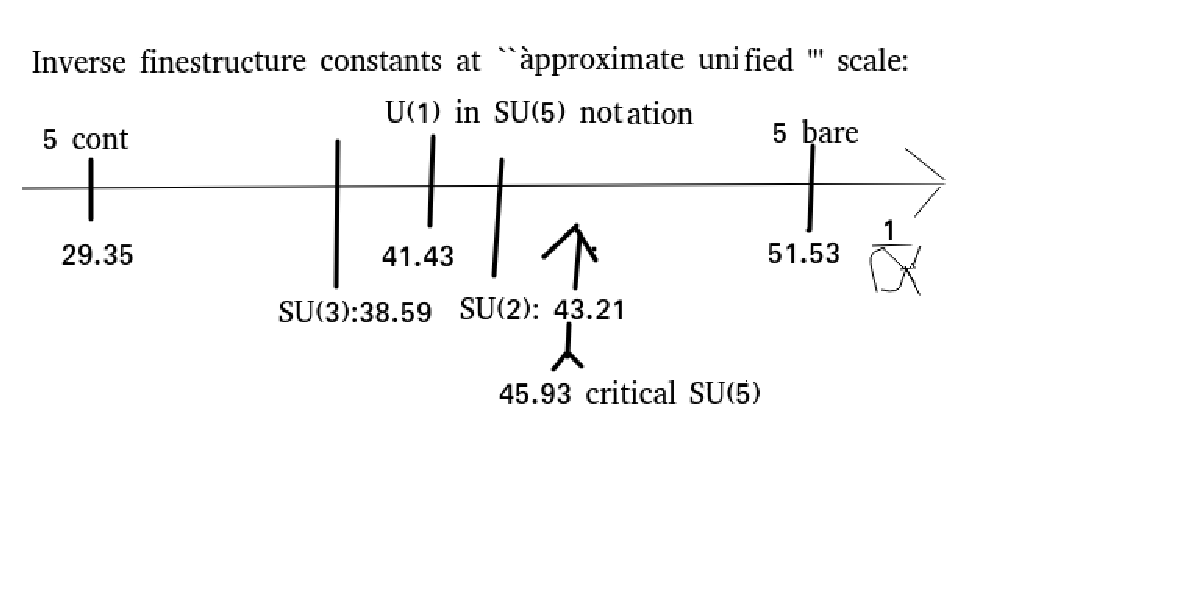}

    \section{What Says Our Result About Original Variables?}
    Our remarkable result is that at the ``unified scale '' $\mu_u$ for
    our {\em approximate } $SU(5)$ the  difference between say
    $\frac{1}{\alpha_2(\mu_u) }$ and $\frac{1}{\alpha_3(\mu_u)}$
      is just the number of
      families $N_{gen}$ {\em times the `unit''} $\frac{\pi}{2}$. 
      It is so to speak the deviation from proper $SU(5)$ symmetry, which seems
      remarkably to be an integer - the number of families - times the
      ``unit''$ \frac{\pi}{2}$, which denotes the amount of shift in
      an inverse $\alpha$ per unit of quantum fluctuations in the lattice
      theory of the theory in question.

      For testing and for illustrating, that there is truly a content
      in our prediction, we want now to rewrite this result in terms of the
      $M_Z$-scale quantities:

      Let us begin to write down the difference, that should have the
      remarkable value $N_{gen}*\frac{\pi}{2}$ (where $N_{gen}$ is the
      number of families):
      \begin{eqnarray}
        \frac{1}{\alpha_2(\mu_u)}-\frac{1}{\alpha_3(\mu_u)}&=&
        \frac{1}{\alpha_2(M_Z)}-\frac{1}{\alpha_3(M_Z)}-\frac{b_2-b_3}{2\pi}
        \ln \frac{\mu_u}{M_Z},\nonumber
        \end{eqnarray}
      where now
      \begin{eqnarray}
        \ln \frac{\mu_u}{M_Z}&=&
        \frac{1/\alpha_2(M_Z) -5/3*1/\alpha_{1\; SU(5)}(M_Z) + 2/3*1/\alpha_3(M_Z)}
             {\frac{b_2-5/3*b_1+2/3*b_3}{2\pi}}\nonumber\\
             \hbox{so that }&&\nonumber\\
             \frac{1}{\alpha_2(\mu_u)} -\frac{1}{\alpha_3(\mu_u)}&=&
             \frac{1}{\alpha_2(M_z)} -\frac{1}{\alpha_3(M_Z)}-\nonumber\\
             && -\frac{b_2-b_3}{b_2-5/3*b_1 +2/3b_3}*\nonumber\\
             &&*(1/\alpha_2(M_Z)
             -5/3*1/\alpha_{1\; SU(5)}(M_Z) + 2/3*1/\alpha_3(M_Z))\nonumber
             \label{ptg}.
             \nonumber
      \end{eqnarray}

      Here the ratio of the $b_i$'s becomes:
      \begin{eqnarray}
        \frac{b_2-b_3}{b_2 -5/3 *b_1+2/3*b_3}&=& \frac{-19/6-(-7)}{-19/6
          -5/3*(41/10)+2/3*(-7)}\nonumber\\
        &=& \frac{-190 +420}{-190 -5/3(+246)+2/3*(-420}\nonumber\\
        &=& \frac{-570 +1260}{-570-1230-840}\\
        &=& \frac{690}{2640}\\
        &=& \frac{23}{88}\\
        Numerically&& \nonumber \\
        \frac{-3.166+7.000}{-3.166-5/3*4.100-2/3*7.000}\\
        &=& \frac{3.834}{-3.166-6.8333-4.666}\nonumber\\
        &=&\frac{3.834}{-14.6653} \\
        &=& -0.26143 \hbox{(agree with} \frac{23}{88} ) \\
      \end{eqnarray}

      Our difference is
      \begin{eqnarray}
        \frac{1}{\alpha_2(\mu_u)} -\frac{1}{\alpha_3(\mu_u)}&=&
        (\frac{111}{88\alpha_2 }  - \frac{115}{264 \alpha_{1\; SU(5)}}-\frac{218}
             {264
          \alpha_3})|_{M_Z}\nonumber\\
        &=& \left ((\frac{111}{88}+3/5*\frac{115}{264})
             \frac{1}{\alpha_{EM}}sin^2\Theta
             -
             3/5*\frac{115}{264}\frac{1}{\alpha_{EM}}
             -
        \frac{218}{264}*\frac{1}{\alpha_3}
        \right )|_{M_Z}\nonumber\\
        &=&\left ( \frac{1}{\alpha_{EM}}*(\frac{333
          +69}{264}*sin^2\Theta
        -\frac{115}{264})-\frac{218}{264}*\frac{1}{\alpha_3}\right )|_{M_Z}
        \nonumber\\
         &=&\left ( \frac{1}{\alpha_{EM}}*(\frac{402}{264}*sin^2\Theta
        -3/5\frac{115}{264})-\frac{218}{264}*\frac{1}{\alpha_3}\right )|_{M_Z}
        \nonumber\\
        &=& \left ( \frac{1}{\alpha_{EM}}*(\frac{201}{132}sin^2\Theta
        -\frac{69}{264})-\frac{218}{264}*\frac{1}{\alpha_3}\right )|_{M_Z}
        \nonumber
      \end{eqnarray}

        \subsection{Calculating Our difference $\frac{1}{\alpha_2(\mu_u)}
          - \frac{1}{\alpha_3(\mu_u)}$ from $M_Z$ scale Data.}

        Let us use
        \begin{eqnarray}
          \frac{1}{\alpha_{EM}(M_Z)} &=&
          127.916\pm 0.015\\
          sin^2\Theta &=&
          0.23116\pm 0.00013  \\
          \alpha_3(M_Z)&=&
          0.1184\pm 0.0007     
        \end{eqnarray}

        Then our difference becomes:
        \begin{eqnarray}
          ``difference''&=& \frac{1}{\alpha_2(\mu_U)} -
          \frac{1}{\alpha_3(\mu_U)}\\
          &=&  \left ( \frac{1}{\alpha_{EM}}*(\frac{201}{132}sin^2\Theta
          -\frac{69}{264})-\frac{109}{132}*\frac{1}{\alpha_3}\right )|_{M_Z}
          \nonumber\\
          &=&
         {\bf (} (127.916 \pm 0.015)*(\frac{201}{132}*(0.23116\pm 0.00013)
          -\frac{69}{264})-\nonumber\\
          &-&\frac{109}{132}*\frac{1}{0.1184\pm 0.0007}
          {\bf )}\\
          &=& 4.6187\pm 0.0014 \hbox{(from $\alpha_{EM}$)} \pm 0.025 \hbox{(from
            $sin^2\theta$)} \pm 0.041 \hbox{(from $\alpha_3$)}\nonumber\\
          &\stackrel{?}{=} & 3*\pi/2 = 4.7124\\
          deviation&=& 0.0937 \pm 0.046\\
          deviation &is\;  about & 2 s.d. 
        \end{eqnarray}

        If you would like to blame all our deviation on the strong $\alpha_3$,
        we would get, that in stead of the used 0.1184 a number 2.3 standard
        deviation
        higher, meaning the replacement,
        \begin{eqnarray}
          \alpha_3(M_Z) = 0.1184 \pm 0.0007 &\rightarrow & 0.1200\\
          \hbox{A strengthning by } 0.0016 & meaning& 2.3 s.d.
          \end{eqnarray}
            \section{Alternative Way of Calculating}
            As an alternative - or check - we could impose our predicted
            values for the differences of the inverse fine structure constants
            and in that way obtain a ``unification scale'' $\mu_U$. If our
            model is right then fitting the ``unification scale'' to the
            different differences between the three inverse fine structure
            constants in the standard model should lead to the {\em same}
            ``unification scale''.

            Let us as the first example take the difference
            $\frac{1}{\alpha_2(\mu)} -\frac{1}{\alpha_{1\; SU(5)}}$. At the
              $M_Z$ scale we have:
            \begin{eqnarray}
              difference_{21}&=&
                \left (  \frac{1}{\alpha_2}-\frac{1}{\alpha_{1\; SU(5)}}\right
                )|_{M_Z}\\
                &=&\left ( \frac{1}{\alpha_{EM}(M_Z)}*sin^2\Theta_W
                -\frac{3}{5}*
                \frac{1}{\alpha_{EM}}*cos^2\Theta_W\right )|_{M_Z}\\
                &=& (-\frac{3}{5} + \frac{8}{5}sin^2\Theta_W)*
                \frac{1}{\alpha_{EM}}\\
                &=&(-3/5 + 8/5 *(0.23116\pm 0.00013))*(127.916\pm 0.015)\\
                &=& (-0.230144\pm 0.00020)*(127.916\pm 0.015)\\
                &=& -29.4390999\pm 0.02
              \end{eqnarray}
              The slope by renorm group of this difference is
\begin{eqnarray}
  \frac{b_2- b_{1 \; SU(5)}}{2\pi}& =&\frac{-19/6 -41/10}{2\pi}\\
  &=& \frac{-95 -123}{2\pi *30}\\
  &=& \frac{-218}{60\pi}\\
  &=& 1.156526
\end{eqnarray}

Now our model - with its quantum fluctuations - says that at the ``unified
scale'' of interest in our model the difference, 2 to 1, shall have run to
\begin{eqnarray}
\hbox{``difference''}_{2 \; to \; 1}&=& (3 -13/5)*3 *\frac{\pi}{2}\\
  &=& 2/5*3*\pi/2\\
  &=& 1.884954.
  \end{eqnarray}
So the ratio of the our ``unified scale'' to the $M_Z$-scale has the logarithm
\begin{eqnarray}
  \ln(\frac{\mu_U}{M_Z}) &=& \frac{1.884954 -(-29.4390999) }{1.156526}\\
  &=& \frac{31.32405}{1.156526}\\
  &=& 27.084608474 \pm 0.02\\
  \frac{\mu_U}{M_Z}&=& 5.79023*10^{11}\\
  \hbox{``unifying scale'' $\mu_u$} &=& 5.27997*10^{13}GeV \pm 10^{12}GeV
  \end{eqnarray}
We earlier got by different calculation
$27.0566$ giving with $M_Z=91.1876 GeV$ that the
``unifying scale'' $5.134*10^{12}GeV$.

Note that the difference between the two different fits to the
$\ln(\frac{\mu_U}{M_Z})$ deviate by just 0.03 while the predicted
quantity 1.88 we used would give rise to a contribution to this logarith
of the order of 1.6, which is more than 50 times larger.So we can claim that
the prediction works well to about 2\% accuracy.

In the table \ref{tableback} we have collected similar calculations for the
other two differences, too.


\subsection{The 2 Minus 3 Case}
We could estimate the same ``unification scale'' $\mu_U$ logarithm
$\ln(\frac{\mu_U}{M_Z})$ similarly using another difference predicted as
e.g. $(\frac{1}{\alpha_2}-\frac{1}{\alpha_3})|_{\mu_U}=1*3*\pi/2
=4.712385$.

At the $M_Z$-scale we have
%
\begin{eqnarray}
  \frac{1}{\alpha_2(M_Z) } - \frac{1}{\alpha_3(M_Z)}&=& \frac{1}{\alpha_{EM}(M_Z)}
  *sin^2\Theta_W - \frac{1}{\alpha_3(M_Z)}\nonumber\\
  &=& (127.916\pm 0.015) *(0.23116\pm 0.00015)-\frac{1}{
    0.1184 \pm 0.0001}\\
  &=& (29.5691\pm0.003)-
  8.4459\pm 0.01\\
  &=&
  21.1232\pm 0.01,\\
  \hbox{but at $\mu_U$ we predict : } (\frac{1}{\alpha_2}-
  \frac{1}{\alpha_3})|_{\mu_U}&=& 1*3*\pi/2\\
  &=& 4.712385\\
  \hbox{Running needed: } ``run \; need'' &=&
  21.1232-4.72385\\
  &=&
  16.3993
\end{eqnarray}
and this difference run with the renorm group by the rate
\begin{eqnarray}
  \frac{d(1/\alpha_2 -1/\alpha_3)}{d\mu}&=& \frac{19/6 -7}{2\pi}\\
  &=&\frac{-23}{6*2\pi}\\
  &=& -0.610094.
  \end{eqnarray}
So the natural logarith of ratio
\begin{eqnarray}
  \ln(\frac{\mu_U}{M_Z}) &=&\frac{
  16.3993}{0.610094} \\
  &=&
  26.8800
  \pm 0.02
\end{eqnarray}
This is to be compared with the  $27.085 \pm 0.02$ from above, and
deviate by about $0.20$, which with an uncertainty for the difference
between the two numbers put to $0.03$ would be $7 s.d.$. But note that
even with this not so impressive number of standard deviations the
deviation of $0.20$ is compared to the number $4.7123/0.6101 = 7.725$
corresponding to our prediction of the value at the unified scale,
about 30 times as small. So our theory works in that sense to $3\%$ accuracy.

\subsection{Superfluous Case Difference 1 to 3}
Although it is just related to the two foregoing let us explicitely for check
also calculate what the our requirement for the difference 1to 3 means
%
\begin{eqnarray}
  \frac{1}{\alpha_{1\; SU(5)}(M_z)} -\frac{1}{\alpha_3(M_Z)} &=&
  \frac{1}{\alpha_{EM}(M_Z)}*cos\Theta_W(M_Z)*3/5 -\frac{1}{\alpha_3(M_Z)}\\
  &=& (127.916\pm 0.015) *(1-0.23116\pm 0.00013) *3/5 -1/(0.1184\pm0.0007)
  \nonumber\\
   &=& 59.0082\pm 0.02 -(8.4459\pm 0.7\%)\\
  &=&50.5623\pm 0.021.
\end{eqnarray}
Then
\begin{eqnarray}
  \ln(\frac{\mu_U}{M_Z}) &=& \frac{(50.5623-3/5 *3*\pi/2)*2\pi}{41/10+7}
  \nonumber\\
&=& \frac{47.7349*2\pi}{111/10}\\
&=& 27.0204 \pm 0.01
\end{eqnarray}

\subsection{Table }

\begin{table}[!]
\caption{\label{tableback}Table of results for three - not indenpent -
  ways of using
  the by us predicted differences between the running inverse fine
  structure constants at ``unfied scale in our model'' (which is the
  scale at which the three running differences should be equal to the
  numbers in line 3 or 4. These predictions are to be fullfilled at this
  ``unified scale'' which using each of the three differences is written
  in line 8, and the success of our model is really that these three
  numbers agree. They deviate from their average $27.04$ by the numbers of
  standard deviations (s.d.) given in line 12. The ``small'' deviations
   agree within accuracy. But more important is to compare these
  deviations from the common average to the ratios given in line 9. which
  should be the contribution from our prediction numbers translated into
  the numbers in $\ln(\frac{\mu_U}{M_Z}$, which we gave in line 8. Here it
  turns out that the deviations from the average of the three numbers
  as written in line 12 in terms of standard deviations, when compared to these
  predictions divided by the running rate are relatively small as seen
  in line 11. In fact these numbers in line 11 are at most of the order of
  1/20, while the two smaller ones of them are only of the order of 1/70.
  This means that our prediction values turned out correctly to better than
  5\%. A similar conclusion would be reached by instead of the average of
  the three 
  $\ln(\frac{\mu_U}{M_Z})$ fits using the value of the
  $\ln(\frac{\mu_U}{M_Z})$ fitted by directly insisting on the ratio of the
  differences of th the inverse fine structure constants being the one we
  require. This insisting on the ratio of the differences directly lead to
  $27.03$, which is only deviating by 0.01 from the average here in the table
  which was
  $27.04$,when wieighting with uncertaitties wereused in evaluating the
  average (the naive average is rather 27.00) . The difference o.o3 is only 1.5 s.d. and quite small compared
  the to the predictions corresponding shifts in the $\ln(\frac{\mu_U}{M_Z})$
  as seen in line 9.or 10. Again this fact ensures that our agreement although
  not perfect (yet) is remarkably good.}
  \begin{adjustbox}{max width =\textwidth}
\begin{tabular}{|c|c|c|c|c|c|c|c|}
  \hline
1. & &$1/\alpha_2 -1/\alpha_{1\; SU5}$&$\pm$ &$1/\alpha_2 -1/\alpha_3$
  &$\pm$&$1/\alpha_{1\; SU5}- 1/\alpha_3$&$\pm$\\
2.&$dif|_{M_Z}$ &$-29.4390$&$
0.03$&$21.1232$&$
0.05$&$50.5623$&$
0.05$\\
  3.&$dif|_{\mu_U \; pred.}$&$2/5*3*\pi/2$&&$1*3*\pi/2$&&$3/5*3*\pi/2$&\\
  4.&&=1.88495&&=4.71239&&=2.82743&\\
  5. &dist to run&$31.32405$   &&$-16.3993$&&$-47.7349$&\\
 6. &Run rate & $\frac{19/6+41/10}{2\pi}$&&$\frac{19/6-7}{2\pi}$&&
  $\frac{-41/10 -7}{2\pi}$&\\
  7.&&=1.156526&&=-0.6101&&=-1.76662&\\
  8.&$\ln(\frac{\mu_U}{M_Z})$&27.0846 &
  0.03& 26.8797 &
  0.1& 27.02046&
  0.03\\
  & as av.+dev.&27.04+0.0446&& 27.04-0.1203&&27.04-0.01954&\\ 
  \hline
  9.&$|\frac{dif|_{\mu_U \; pred.}}{\hbox{Run rate}}|$&$|\frac{1.88}{1.15}|$&
    &$|\frac{4.71}{-0.610}|$&&$|\frac{2.827}{-1.7666}|$&\\
  10.&&=1.629&&=7.724&&= 1.6005&\\
  11.& rel.dev.&0.052&&-0.015&&0.011&\\
  12.& $\ln(\frac{\mu_U}{M_Z})$
  & 1.5 && 1.6 && 0.6&\\
&s.d.f. av.&&&&&&\\
  13. &d. fr. $27.03$&0.05&&-0.15 && -0.01&\\
  \hline
\end{tabular}
\end{adjustbox}
\end{table}

By accident the average of the three values for $\ln(\frac{\mu_U}{M_Z})$ turns
out to be exactly $27.00$ within our uncertainty. The 11th line in the
table gives the deviation from this average relative to the part of the
$\ln(\frac{\mu_U}{M_Z})$, which is due to our prediction value, so it give
the order of magnitude of the failure of our prediction relatively. Remark,
that even the biggest of these three deviation measures relative to our
predictions is {\bf only 0.052} meaning that even this deviation is only
so well fitting by accident in one out of 24 cases. 

The $\ln(\frac{\mu_U}{M_Z}) =27.00$ correspond to that the
``unification scale in our model''
\begin{eqnarray}
  \frac{\mu_U}{M_Z} &=& 5.32 *10^{11}\\
 \hbox{ and } \mu_U &=& 4.85*10^{13} GeV
\end{eqnarray}

\section{ Critical Coupling}
\label{ps10}
Now we have without using but the lattice theory philosophy - see the old
works and \cite{Ferrucchio}, also connection to our several phase
speculations \cite{sf} - reached to
        an understanding in our picture of the deviations from the $SU(5)$
        symmetry. It would of course be natural first to look for, if
        the unifying coupling should be the critical one for $SU(5)$
        corrected of course for the factor that is the number of families.
        This is though not at all obviously the correct thing to do
        in our philosophy, because we have in the philosophy of the present
        article no true $SU(5)$ theory. It is only approximate, but lacks
        half of the degrees of freedom. nevertheless let us for first
        orientation look for comparing the expression for the $SU(5)$
        critical coupling given by Laperashvili, Ryzhikh, and Das
        \cite{LD, LRN}
        \begin{eqnarray}
          \alpha_{N \; crit}^{-1} &=& \frac{N}{2}\sqrt{\frac{N+1}{N-1}}
          \alpha_{U(1) crit}^{-1}
          \end{eqnarray}
        where we for the critical $U(1)$ coupling take the lattice value
        for Wilson and Villain actions:
        \begin{eqnarray}
          \alpha_{crit}^{lat} &\approx &0.2\pm 0.015.
        \end{eqnarray}
        This gives
        \begin{eqnarray}
          \alpha_{5 \; crit}^{-1} &=& 0.2^{-1}*5/2*\sqrt{3/2} = 5*5/2 *1.2247\\
          &=& 15.309 \label{cr}
          \end{eqnarray}

        With the family factor $N_{gen}=3$ this would let us expect
        $15.309*3 = 45.927$ to be compared with the estimates from data above.
        (Presumably) the value to compare with is the $51.5$ for the
        unified coupling not corrected by the quantum fluctuations, which we
        considered so much in this paper. Now we must remember, that
        the $U(1)$-critical coupling was $0.2\pm 0,015$ meaning $7.5 \%$
        uncertainty. These $7.5\%$ means $\pm 3.45$ for the $46$ we predict.
        So the ``experimental'' $51.5$ from our fit is only off by
        $\frac{5.5}{3.45} = 1.6$ s.d.. If there is an uncertainty 
        in the critical coupling formula, we used, in addition to the one from
        the uncertainty in the critical coupling for $U(1)$, then
        the deviation in standard deviations will be even smaller than the
        1.6.

        So formally we must count the hypotesis, that indeed the critical
        inverse unified finestructure constant should be just 3 times the
        critical one, is very successfull! One should have in mind, that in
        reality the ``the critial finstructure constant'' is not quite well
        defined, because it depends on the details of the lattice theory.

        If we accept this agreement, we can say, that we fitted all three
        Standar
        Model fine structure constants with {\bf only the unification scale},
        i.e. {\bf one} paramter. The unification value of the fine structure
        constant for the $SU(5)$ was determined by the ``critcallity''.

        Actually we shall even below in section \ref{scale} claim that
        we can relate the approximate unification scale - the lacking
        parameter to predict at this stage in the article - to the
        top-mass and the Planck scale,so that at the end we shall have
        predicted all three parameters.
        
        \subsection{More Thoughts on the Critical Coupling, Unified Coupling}

        Thinking a bit deeper: We should really not take a formula for the
        $SU(5)$ critical coupling without correction, because we have been
        claiming all through the article that in our model the $SU(5)$
        symmetry and all its degrees of freedom do not exist. Rather we
        should look
        for correcting the number for the critical  $\alpha_5$
        to the critial coupling for the lattice standard Model group coupling:

        Very crudely we think of the critial coupling for groups like the
        ones we look at to be the transition between two phases described as
        \begin{itemize}
        \item{1.} An essentially classical phase, wherein the coupling is so
          weak - i.e. $1/\alpha$ so large, that at the scale we consider
          (the lattice links cale)
          all the plaquette variables are so close to unity, that the
          quantum effects can be considered just perturbations, but that
          basically we have the classical theory working.
        \item{2.} A ``confined'' phase, in which we rather have that to
          first approximation the plaquette variables are distributed
          uniformly all over the group volume,as the Haar measure, we
          could say. Of course it will be still be  more likely to
          find the plaquette variables closer to the unit element
          in the group until the inverse coupling $1/\alpha$ reaches zero.
          But now it is the variation of the probability density over the group
          that is the ``small'' perturbation.
        \end{itemize}

        If the standard model group lie as a {\bf dence network} inside the
        $SU(5)$ in the ${\bf 5}$-plet vector representation space, then the
        a bit smeared volume of the standard model group would be
        similar to that of $SU(5)$ proper, and the value of the (inverse)
        fine structure constant, at which one or the other one of the two
        approximations above will shift their dominance (i.e. the
        critical value), will be (roughly) the same as for full $SU(5)$.
        But of course the denceness of the net formed by the
        standard model group is not perfect, and thus it will require that
        one goes to a somewhat stronger coupling (i.e. smaller inverse
        $1/\alpha$) to give the `` confinement phase'' enough weight in the
        partition function to (barely) compete with the ``classical phase''.
        So thus we expect
        \begin{eqnarray}
          \frac{1}{\alpha_{SMG\; crit} } &\le&
          \frac{1}{\alpha_{5 \; crit}}\hbox{{\bf but only a bit}}.
          \label{inequality}
        \end{eqnarray}

        But now we have - to be fair - to remember that the standard model
        group, never had the quantum fluctuating degrees of freedom, which the
        full $SU(5)$ lattice gauge theory has. It lacks at least the 12
        degrees of freedom, we refered to by $H_{int}$ in our calculation. So
        going
        from the standard model ``total'' coupling, if such a thing existed,
        to the various subgroups $SU(2),\;  SU(3),\;  and \; U(1)$ would not
        corresond to taking away so many fluctuations as, if one went from
        the full $SU(5)$. So the critical $\frac{1}{\alpha_{crit \; SMG}}$
        should not be identified with the above fitted
        $\frac{1}{\alpha_{5\; bare}}$, but rather with an inverse fine
        structure constant of a type, that shall not have had it fluctuations
        in the set $H_{int}$ type ones removed,as we did in our formalism
        when constructing this ``bare'' inverse $SU(5)$ fine structure
        constant. 
        So what we should rather identify as the implimentation of the critical
        coupling assumption, is to say, that a ``fitted''
        $\frac{1}{\alpha_{SMG}}$
        is the one you get by not counting that the refered to by $H_{int}$
        modes be included, but only the other ones, is to be identified by
        the 3 *$\frac{1}{\alpha_{smg\; crit}}$ which by (\ref{inequality}) is
        - actually only a bit - smaller than $\frac{1}{\alpha_{5 \; bare \; crit}}$.
        The ``fitted'' quantity   $\frac{1}{\alpha_{SMG}}$ comes actually very
        close to be an average of the three inverse fine structure constants
        from the standard model,which is rather expected, since it is the
        standard model genuine gauge group. Then if the dence net-work
        with which the standard model group $G_{SMG}$ covers the $SU(5)$, 
        there will only be little difference between the two sides in
        (\ref{inequality}) and we now expect, that the average of the
        three standard model group inverse fine structure constants at
        ``our unification scale'' say essentially being $\frac{1}{\alpha_{SMG}}$
        shall be a bit smaller than the critical SMG inverse fine structure
        constant times the 3, which again is just a bit smaller than the
        3 times the critical inverse fine structure constant for $SU(5)$:
        \begin{eqnarray}
          41.34& =&\frac{1}{\alpha_{1\; SU(5)}(\mu_U)}\hbox{(taken as average
            $1/\alpha_i$)}
          \\
          &\approx& \frac{1}{\alpha_{SMG}(\mu_U)}\\
              &=& 3*\frac{1}{\alpha_{SMG\; ccrit}}\\
              &\stackrel{<}{\hbox{a bit}}&3* \frac{1}{\alpha_{5\; bare\; crit}}\\
              &=& 3* 15.3=45.9
          \end{eqnarray}

        
        \section{Crude Second Order Calculation}
        \label{ps11}
        We did in principle the above calculations only up to
        first order approximation in a preturbative scheme, in which the
        0th order approximation is the exact $SU(5)$, wherein all three
        standard fine structure constants are equal to each other, and the
        first order approximation is the one, in which our corrections are
        considered small of first order, so that the squares of the corrections
        can be considered negligible. The numerical order of the first order
        quantities are
        \begin{eqnarray}
          \hbox{``first order size''} &\approx & \frac{\alpha}{ 1 \hbox{ or }
            3*\pi/2}\\
          &\approx& 1/10.
          \end{eqnarray}
        One unneccesary ignorance of second order terms, which are expected to
        be of the order $(1/10)^2$ times the main term is, that we above let the
        $\alpha$ appearing as a factor in the $<H^2>$'s cancel with the
        $1/\alpha$ whichever among the $1/\alpha_i$'s we meet. Actually
        it was tempting to think, that by using this lucky trick
        of getting rid of the parameters in the estimate of our
        corrections, we were likely actually to get a better result
        with respect to agreeing for our calculations. Once we look for
        accuracies of the second order, there may be more corrections, such as
        the $H$ distribtuion being not even just Gaussian, and the whole
        program of doing the second order deserves a further article. Here we
        shall only make a very crude attempt to estimate the effect of
        seeing what $\alpha$ (among the three one say) comes into which of the
        fluctuations $<H^2>$. We shall make the assumption that the $\alpha$
        to be used for an $H_i$ where it is the fuctuation in one of the basis
        vectors for the subgroup $i$ of the $SU(5)$, is $\alpha_i$.
        Then we see from table \ref{tabcorr} that we have had relativley good
        luck by letting the two $\alpha$'s cancel each other, because the
        mostly contributing $<H_i^2>$ to the correction for the inverse
        fine structure constant $1/\alpha_j$, for the standard model
        subgroup denoted $j$ is actually mostly $i=j$ itself. In fact e.g.
        according to the table \ref{tabcorr} the correction to
        \begin{eqnarray}
          \hbox{Fraction of SU(2)-inverse coupling not $H_2$ } \frac{3/10}{3/2}
          &=& \frac{1}{10}\\
          \hbox{Fraction of SU(3)-inverse coupling not in $H_3$ }
          \frac{2/15}{8/3} &=& \frac{1}{20}
        \end{eqnarray}
        For the $U(1)$ inverse fine structure constant the dominant
        contributuion to the corrections comes from the two nonabelian groups,
        i.e. from $H_2$ and $H_3$, but it has a bigger number from
        the $H_1$ than any of the other two groups, namely $7/30$.
        But since the $U(1)$ coupling correction is so mixed, to take
        all the same $\alpha$ is not so bad.

        In any case it looks that it is only about 1/10 of the correction for
        the SU(2) coupling and 1/20 for the SU(3) coupling, that would be
        changed by being a bit more carefull with which $\alpha$ to use.
        The change to the more correct $\alpha$ to use would thus increase
        difference $1/\alpha_2-1/\alpha_3$ percentwise
        by
        \begin{eqnarray}
          \hbox{Decrease of $1/\alpha_2 -1/\alpha_3$ } &=&
          \frac{1/10 +1/20}{2} *4.7/2/40\\
          &=& 3/40 *0.06\\
          &=& 0.045 \hbox{ relatively}
        \end{eqnarray}
        This is now to be compared with the deviation of
        of the $3*\pi/2=4.712385$ from  the number in (\ref{d23}) which is
        $4.62$ and thus smaller than or prediction  $3*\pi/2=4.712385$ by
        $0.09$ which relatively is $0.0190$. This agrees only modulo a
        factor 2.
        
        The observed by renorm group developping the fine structure constants
        to the ``our unification scale'' defined from the ratios of the
        two independent differences of inverse couplings to be 2:3 was 4.62,
        i.e. smaller than the theoretical 4.71, but now the effect of
        pushing the inverse finstructure constants predicted down from their
        starting point in the SU(5)-symmetric limit $1/\alpha_{5\; naive}$
        is getting increased for the $SU(2)$-inverse fine structure constant,
        because for that the changed $H_1$ contribution is getting increased
        by our second order correction because the $\alpha_{1\; SU(5)}$
        is correctly stonger than what we used at first. For the $1/\alpha_3$
        oppsitely the $1/\alpha_{1\; SU(5)}$ is above the $1/\alpha_3$ at the
        ``our unification '' so that for the $1/\alpha_3$ the $H_1$
        contribution corresponds to a weaker $1/\alpha_{1\; SU(5)}$ thus
        giving a lower suppression compared to naive inverse SU(5) coupling,
        $1/\alpha_{5\; naive}$. Thus the theoretical $4.712385$ should be
        deminshed - since the 3- inverse coupling goes up by the correction
        and 2-inverse coupling down - relatively by the $0.045$. But that would
        bring the theoretical number to $4.50$, close to the $4.62$.

        The deviation from the only to first order result of the
        number gotten by fitting is of the order of magnitude of the
        second order estimate. So it is important to estimate this second order
        approach more carefully.
        \section{Speculative Relation to Planck Scale}
        \label{scale}

        A major problem and surprize comming, if one takes our suggestion of
        the truly existing lattice at the approximate or ours unification scale
        $\mu_U$ = $5.18*10^{13} GeV$ seriously is, that it suggests a
        ``fundamental''
        scale {\bf quite different from the Planck scale}. To seek a way
        out of this problem we propose to think of a {\bf fluctuating
          lattice even in size of the lattice constant} in the sense that
        we speculate, that the general theory of relativity is still
        perturbatively treatable and rather well understood already
        - so that no completely speculated quatum gravity theory is
        needed at the $\mu_U$ scale - so that the whole lattice structure
        must be in a quantum superpostion state invariant under the
        reparametrization group from the general relativity. That is to say,
        with the philosophy, that there is very big quantum fluctuations in
        the gauge and taking the diffeomorphism of reparametrization
        symmetry as the gauge symmetry of general relativity, we must take
        it that the world is in a superposition of all the possible deformations
        of the lattice - needed for our model for the approximate GUT $SU(5)$ -
        achieved by reparametrizations. That is to say, that in a typical
        component in this superpostion somewhere we find a very small
        lattice constant and somewherewe find a very big one, so that lattice
        cannot be exactly a Wilson one e.g.. But locally it could still
        be close to a Wilson lattice. Then of course the lattice constant
        value suggested by our parameter $\mu_U$ as lattice constant
        $a \approx 1/\mu_U$ could only be true in an average sense:
        \begin{eqnarray}
          \mu_U &=& {\bf Average} \frac{1}{a},
          \end{eqnarray}
        where $a$ is some local, or may be better single link, lattice constant,
        i.e. length of the link in the metric of the general relativity,
        which should still be perturbatively treatable in the range
        around $1/a \approx 5.18*10^{13} GeV$ (which is a small energy relative
        to the Planck scale).

        So the physical model, in which we developped our more primitive
        lattice model, is in the rest of the article further developped
        into
        {\bf
          some presumably more chaotic lattice theory (a kind of glass), in
          which the degree of fineness varies from region to region and you
          find links of all possible sizes, and at least approximate
          diffeomorphism invariant structure of the lattice.} It is of course
        only approximately diffeomrphisminvariant by being in superposition
        of having different fineness of the lattice at any place.
        From the approximate diffeomorphism invariant structure of the
        lattice model in this section we cannot avoid, that the density
        of links of the length around $a$ has to vary approximately like
        \begin{eqnarray}
          \hbox{``density''}(\ln(a)) d\ln(a)&=&
          P(\ln(a)<\ln \hbox{``link length''}<\ln( a) +d\ln(a))\nonumber\\
          &=& a^{-4}d\ln(a),
          \end{eqnarray}
        where  $P(\ln(a)<\ln \hbox{``link length''}<\ln( a) +d\ln(a))$
        is the probablity of finding a random link taken out of our
        ``chaotic lattice'' within
        the scale in logartihm from $\ln(a) < \ln(\hbox{lattice constant })<
        \ln(a)+d\ln(a)$. A similar distribtion of the sizes of the plaquettes
        found in the ``chaotic lattice'' of this section,  would also
        have a factor in the density going as the fourth power of the inverse
        plaquette side size.

        There is actually a
        divergence problem with this
        ``chaotic lattice'' as we speculate it: If indeed this density
        distribution should be fully true the probablity of finding links
        of a specific order of magnitude would need to be zero and all
        the contribtuion would come from infinitely small links or infinitely
        long link. So we have to imagine that there finally must be some
        cut offs for very long - not so important - and for very short links
        at least.


        To have approximate diffeomorphism symmetry and thus also
        approximate scale-invariance we should have at most a very slowly
        varying weight factor depending on the logarithm of the say link
        length
        - but only very weakly breaing  the scale symmetry in the range of
        scales we consider relevant, meaning scales between the Planck scale
        and macroscopic scales.

        But if we scall be concrete wewould propose aGaussian weighting
        as a function of the logarithm of say the link length. Near the peak
        in the Gaussian such a Guassian weighting is only very weakly
        breaking the scaling invariance, but for very large or very small
        scales the Gaussian distribution of the weighting in the logarithm
        is enoermous. But somehow we hope that for very small or very
        big link length we have got the cut off effectively and there are
        anyway so little chanse for the links having that size that it does not
        matter so much.  But I think we need a cut off in this style of being
        smooth for some ``relevant'' region and then very drastically cutting
        off  in the scales ofvery small $a$ (i.e. high energies) becuase
        if we did not have the strong cut off somewhere, then attempting to
        play simultaneous with the extra factor $(1/a)^4$ for the Standard
        model approximate $SU(5)$ and an other extra factor $(1/a)^6$ for
        describeing the generalrelativity Einstein-Hilbert action  would
        unavoidably lead to severe divergensies.

        We could say that the proposed Gaussian as function of the logarithm
        very robust by being able to cut off at the ends large and big scales
        any polynomial extra factor. Then in addition to in this way be able to
        cope with any power extra factor, it can be claimed in the
        appropriate region of scales to be rather flat, so that is not
        at such scales drastically breaking scale symmetry.

        With this very special ``cut off'' assumption, it might be felt needed
        to make at least a little bit of propaganda for it: Once we preferably
        should have had invariance under scalings in size,it is suggested
        that we need a slowly varying weight as function of the logarithm
        of the scale. We also like at the end a robust cut off that can
        cut off anything polynomial say and then an exponential of a smooth
        function
        \begin{eqnarray}
          ``weight'' &\sim & \exp( f(\ln(1/a))
          \end{eqnarray}
        is suggested. But then the Gaussian -which may not be
        so crucial exactly - is gotten by Taylor expanding the function
        $f$ around the maximum, which is of course the most important
        region. One could as propaganda also say, that the cut-off proposed
        represents a weak coupling to the metric tensor of gravity.
        
        Then depending on whether you have a factor $a^{-4}$ as for the
        inverse  fine structure constants or a factor $a^{-6}$ as for the
        gravitational $\kappa$ the weighted maximum in the over scale logarithm
        intergal, will have somewhat different central values, i.e. central
        logarithms of scales.

        These centers of the contributing distrbutions will be the effective
        lattice scales for the different weightings. So we can indeed get the
        $\mu_U$ weighted with $a^{-4}$ and the gravitational scale being the
        central one for weight $a^{-6}$ become different by
        orders of magnitude. If we just at first give a name
        to scale $\mu_0$ which one gets with weight $1$, then in the Taylor
        expansion lowest order approximation the drag shifting will be in the
        ratio 6 : 4, so that
        \begin{eqnarray}
          \ln(\frac{E_{Pl}}{\mu_0}) &=& 6/4\ln(\frac{\mu_U}{\mu_0}).
          \end{eqnarray}
        (whether one shall use the formal Planck constant just made by
        dimensional arguments from the Newton constant $G$ or some reduced
        one with an extra factor $8\pi$ extracted might be discussed, but
        may be just considered an uncertainty)

        \subsection{Averaging over Our ``Chaotic Lattice''}
        When we have some part of the continuum lagrangian like the
        $\frac{2\pi}{\alpha} F_{\mu\nu}F^{\mu\nu}d^4x$, then the contribution
        to it in the lattice theory - our chaotic one or just a usual
        Wilson lattice -  come from individual plaquettes or whatever
        combination of the lattice ingredients, that contribute, but you get
        therefore a bigger contribution the more of these contributing objects
        there are per hypercubic unit volume to the coefficient in the of the
        continuum lagrangian density.

        Actually we can use simple dimensional arguments to see how the average
        of the continuum Lagrangian coefficient comes about:

        For the inverse fine structure constants you simply get a
        contribution to
        the action from each plaquette independent of its size (provided
        you let the $\beta$ weighting the plaquette in the action be the same
        whichever the size of the plaquette, especially with our
        philosophy that it should be critical such $beta$ independent
        of the size is suggested.). So in terms
        of an integral over the logarithm of the inverse size, say $1/a$, of the
        laticce constant or link-length we have
        \begin{eqnarray}
          1/\alpha &\propto& \int (1/a)^4 \hbox{``cut off weight ''}d\ln(1/a)\\
          &\propto& \int(1/a)^3 \hbox{``cut off weight'' }d(1/a)\\
          &\propto& (1/a)^4|_{\hbox{at peak for $(1/a)^4$ *weight}}\\
          \hbox{But gravity, extra $1/a^2$: }
          &&\nonumber\\
          \kappa
          &\propto& \int (1/a)^4*(1/a)^2\hbox{``cut off weight''}d\ln(1/a)\\
          &\propto& \int (1/a)^5\hbox{``cut off weight''}d(1/a)\\
          &\propto& (1/a)^6|_{\hbox{at peak for $(1/a)^6$*weight}}
          \end{eqnarray}
        So we see that we {\bf predict} from the ``chaotic lattice '' model with
        its approximate scale invariance, by an essentially dimensional
        argument, that there shall be different effective lattice scales
        for the Yang Mills theories $\mu_U$, and for gravity. (But it is of
        course dependent on our Gaussian in log in some sense special
        cut off, although it is suggestive.)

        In the figure \ref{latticescales} we illustrate, how we after
        having inserted
        a strong cut off implementing weight get a distribtuion in the
        logarithm $\ln(1/a)$ of the scale with a broad peak, (which we imagine
        Gaussian, in this log, in first approximation).

        The main point is that the dominant or peak value for the distributions
        depend on the exact distribution, and that the one for gravity
        has got an extra factor $(1/a)^2$. For the Standard Model gauge
        couplings this peak scale is only of relevance via the renomalization
        group, while for gravity the very size of the (inverse) coupling
        $\kappa$ (also) depends on the peak value for the (logarithm of)
        $1/a$.

        It should be clarified, that it is only because of some
        ``phenomenologically'' added  `` cut off weight '' factor
        that we at all mannage to get  a peaking distribution instead
        of some nonsence divergent one, just increasing monotomously.
        So the picture we propose is really much dependent on there being
        some cut off of this type, and this cut off has to be considered
        some sort of ``new physics'', even though we escape from
        assuming  many details about it, except that it is smooth in
        the logarithm of the scale and sufficiently strong to cause the
        convergence (preferably exponential in form, but with a low
        coefficient on the function, say $f(\ln(1/a))=
        \hbox{``small number''}* (\ln(1/a) - const.)^2$,   in
        the exponant.).

        Let us now suppose that including this ```new physics'' weight
        there is scale, which we call $\mu_0$ for which the density of
        plaquettes or links counted per {\bf link-size volume} is maximal. Then
        if we do not put the factor $(1/a)^4$ or $(1/a)^6$ on as we did above,
        then the peak of the so to speak just ``weight'' would be
        at $\mu_0$  or we should say $\ln(\mu_0)$, when thinking of the
        plotting with $\ln(1/a)$ along the abscissa as in
        figure \ref{latticescales}.

        Now in the approximation of the ``weight'' distribtuion being
        Gaussian in the logarithmic scale and noticing that the extra factors
        $(1/a)^4$ and $(1/a)^6$ from the logarithmic abscissa point of
        view are linear terms in the exponent
        $4\ln(1/a)$ and $6\ln(1/a)$ which will shift the peak from
        $\ln(\mu_0)$ by amonts proportional torespectively 4 and 6,
        we see that
        \begin{eqnarray}
          \frac{\ln(\frac{E_{Pl}}{\mu_0})}{\ln(\frac{\mu_U}{\mu_0})}
          &=& \frac{6}{4}
          = \frac{3}{2} \label{pred}
          \end{eqnarray}

        \subsection{On the Maximum Before the Powers in $1/a$  Factors}

        In seeking to guess, what to take for the maximum density scale
        $\mu_0$, when no extra factor like the $(1/a)^4$ or $(1/a)^6$, we
        should have in mind that the density of plaquettes in a volume (in
        four space) of size like the plaquette or link is indeed, what we
        called the number of ``layers'', which again were identified
        with the number of families, or at least this density of plaquettes
        in the  range associated with a plaquette is proportional to
        the number of layers.

        Since we identify by our hypotesis the number of layers with
        the number of families, we take the number of layers at different
        scales to reflect the number of families being present
        as fermions with negligible mass at the various scales.
        That is to say, that in the range of scales of the quark and (charged)
        lepton masses we have region of scales where as one goes down
        in energy loose more and more families. With such a philosophy of
        counting only the effectively massless fermions at the scale we may
        - using a table like table \ref{ql} -  extrapolate to a scale with
        maximal number of families and take that as $\mu_0$; we could take it
        close to the mass of the mostmassive quark or lepton, the
        top. Actually as seen in table \ref{tawgr} putting
        $\mu_0 = m_t $ the top quark mass is close to
        make our prediction (\ref{pred}) be satisfied. Fitting to make
        our prediction (\ref{pred}) be exactwould require a slightly higher
        in energy scale $\mu_0$.

        We consider this close to agreement as success for explaining
        theoretically the ``unification scale'' $\mu_U$ of our approximate
        $SU(5)$.
        
        \begin{table}[!]
            \caption{Here we just listed the charged quarks and leptons exposing
          their masses and the natural logarithms of the latter with
          the purpose of very crudely use them to extrapolate to scale $\mu_0$
          at which the number of at that scale effectively massless flavours
          would be maximal. This scale $ \mu_0$ is presumably very close to the
          top-mass, since  just above $m_t$ all the quarks and leptons
          are effectively massless. But how high above we shall expect the
          maximum for the purpose of our lattice remain speculations. \label{ql}}
        \begin{tabular}{|c|c|c|c|}
          \hline
          Name&Mass&$\ln(Mass/GeV)$&Sums etc.\\
          \hline
          Quarks:&&&\\
          \hline
          up&2.16 MeV&-6.137&\\
          down&4.67MeV&-5.367&\\
          strange&93.4MeV&-2.371&\\
          charme&1.27GeV&0.239&\\
          bottom&4.18GeV&1.430&\\
          top&172.5GeV&5.150&\\
sum quraks&&-7.055&-1.176\\
``average''&309 MeV&-1.176&\\
\hline
electron&0.5109989461MeV& -7.055&\\
muon&105.6583745MeV&-2.248&\\
tau&1776.86MeV&0.575&\\
sum leptons& &-9.252&-3.084\\
``average''&45.78 MeV&-3.084&\\
\hline
av. weight 2:1&163 MeV&$-1.812$&\\
\hline
        \end{tabular}
      
        \end{table}

\subsection{Ambiguity of Concept of Planck Energy Scale, Reduced ?} 
        In reduced Planck units, the Planck energy $1.22 *10^{19} GeV$ from
        unreduced Planck units is divided by $\sqrt{8\pi} = 5.01325$ so as to
        get
        \begin{eqnarray}
          E_{Pl\; red} &=& 1.22*10^{19} GeV/ 5.013225\\
          &=& 2.4335 *10^{18}GeV
        \end{eqnarray}

        Now, however, we must ask: what is it that gives us a
        scale in the sence the studies of the running
        couplings tells us? The ratio of the reduced Planck energy
        $2.43*10^{18}GeV$ relative
        to the logarithmically averaged charged lepton masses
        $m_{average}= 163 MeV$ is
        \begin{eqnarray}
          \frac{2.43*10^{18}GeV}{0.163GeV}&=& 1.4930 *10^{19}\\
          \hbox{and has } \ln(\frac{E_{Pl \; red}}{m_{av. ch. fermions}})&=&44.15\\
          \hbox{Further: } \frac{m_Z}{m_{av.ch.fermions}} &=& \frac{91.1876GeV}
               {163MeV}\\
               &=& 559.4\\
               \hbox{and has }\ln(\frac{M_Z}{m_{av.ch.fermions}})&=& 6.327\\
               \hbox{So for ``our'' scale } \ln(\frac{\mu_U}{m_{av.ch.fermions}})
               &=& 27.05+ 6.327\\
               &=& 33.38\\
               \hbox{Thus the ratio }
               \frac{\ln(\frac{E_{Pl \; red}}{m_{av.cg.fermions}})}
                    {\ln(\frac{\mu_U}{m_{av.ch.fermions}})}
               &=&\frac{44.15}{33.38}\\
               &=& 1.323.
          \end{eqnarray}

        Had we not used the reduced Planck energy, but the usual one, we
        would have got the logarithmic distance from the quark and
        charged lepton mass scale to the Planck one $\ln(\sqrt{8\pi})
        = 1.612$ bigger, so that it would go from the 44.15 up to
        44.15 + 1.612 =45.76.
        Then we would get the ratio changed to
        \begin{eqnarray}
          \frac{\ln(\frac{E_{pl}}{m_{av.ch.fermions}})}
               {\ln(\frac{\mu_U}{m_{av.ch.fermions}})} &=&
               \frac{44.15+1.61}{33.38}\\
               &=& \frac{45.76}{33.38}\\
               &=& 1.371
          \end{eqnarray}
        In fact we think, we can argue for, that this latter choice is not the
        correct one, because the $ 8\pi$ or $4\pi$ usually comes from
        the differnce in the coefficient to a Coulomb field and the charge
        appearing in the field theory action. When we have just used the
        fermion masses without any $4\pi$-like correction we associate
        it with the simple relation $m= g_y<\phi>$, while if I would like
        the Yukawa-field around the Higgs particle I would get a $1/(4\pi)$
        factor in. So the simple masses correspond we could say to the
        Yukawa coupling $g_y$ being used for unit, and not the alternative
        $g_y/(4\pi) $. So to speak
        \begin{eqnarray}
          G &\sim& \frac{g_y}{4\pi}\\
          (4\pi \hbox{or} 8\pi )G &\sim & g_y \hbox{ and thus also }m 
        \end{eqnarray}
        This argues for, that the reduced $E_{Pl\; red}$ was the right one
        to use not to introduce  unjustified extra factors.

        We could also have argued that the nice scheme of the Standard Model
        with its gauge fields and three families is spoiled, when going down
        in energy already at the Higgs scale, so that we should not come up
        with this logarithmically averaged fermion masses, but just
        use the very $Z^0$ mass $M_Z$  instead, then our ratio would be a bit
        simpler to compute:
        \begin{eqnarray}
          \frac{\ln(\frac{E_{Pl \; red}}{M_Z})}{\ln(\frac{\mu_U}{M_Z})}
          &=& \frac{-1.612 +\ln(\frac{1.22 *10^{19}GeV}{91.1876 GeV})}{27.05}\\
          &=& \frac{-1.612 +39.43}{27.05}\\
          &=& 1.398
        \end{eqnarray}

       
        \subsection{Table of Combinations}
        The most important outcome of the fluctuating-size-of-links lattice, we
        propose, is that it gives us the possibility of having a Planck scale
        very
        different from the ``unification scale'' and still claim a
        ``fundamental''
        lattice at the unification scale. But we would of course like to
        see, if the
        order of magnitudes are at all thinkable. We therefore in
        figure 11.4
        illustrate how we imagine a smooth Gaussian distribution in the logarith
        of the link length say.

\begin{figure}[h]
          \includegraphics[scale=0.8]{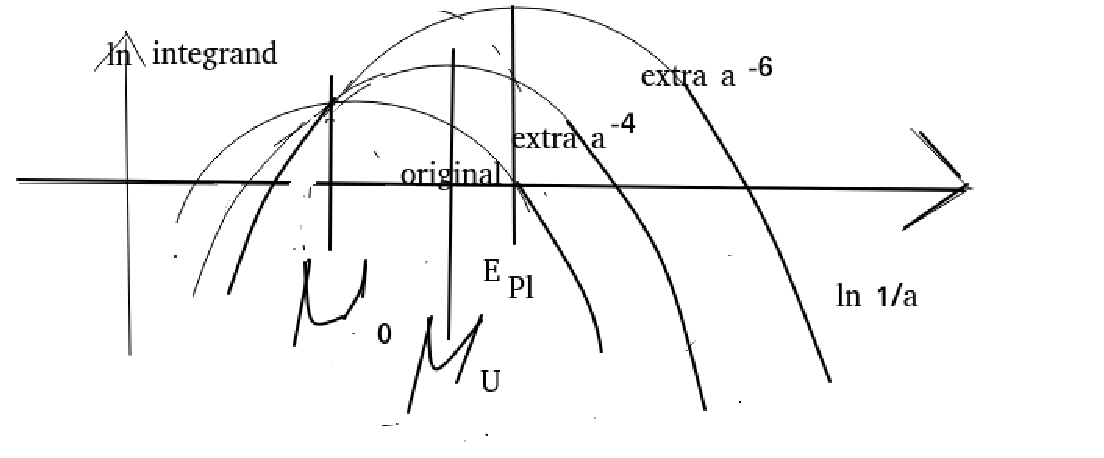}
         
          \caption{As function of the logarithm of the scale - say
            given as energy being the inverse of the link length $1/a$
            we give here the 1) density of links perlinks-length to the fourth,
            2) this density multiplied by $a^{-4}$,and that is the contribution
              to the Lagrangian dentity for Yang Mills theories, 3) the first
              density multiplied by $a^{-6}$, and that is the density of
              contribution to the Einstein Hilbert Lagrangian density.
              See also the text. In our approximation we assume these
              densities to be Gaussian,and with the logarithmic ordinate
              these Gaussians are parabolas pointing downwards. \label{latticescales}}
        \end{figure}

        {\bf Description of figure\ref{latticescales}:}
                      Here the number densities of links
            or of plaquettes, in a small length range of say a percent
            counted or weighted in different ways. The curve
            ``original'' is for counting this number density as the number in
            4-cube of size proportional to the link length range which is being
            counted. In the two other curves the ``original'' density has been
            weighted with respectively the inverse fourth power of the
            link-length $a$ and the sixth power. For all three curves it is the
            logarithm of the density, which is plotted and a Gaussian
            behavior as function of the logarithm of the inverse length of
            the link is assumed as suggestive example. Plotted with
            logarithmic ordinate of course a Gaussian distribution looks like
            a downward pointing parabola, and the three curves are meant
            to be such downward pointing parbolas. It is trivial algebra to see
            that weighting the density counted the ``original'' way by further
            respectively $(1/a)^4$ and $(1/a)^6$, the logarithms of which are
            linear in $\ln(1/a)$, just leads to displacements of the p
            parbolas,
            but leave their shapes the same. For the fine structure constants
           or say our approximate $SU(5)$ it is the total number of plaquettes
            equivalent to the weighting with $(1/a)^4$ that counts, and the
            effective lattice link-size for our approximate $SU(5)$ model
            should thus be the tip of the distribution with the ``extra factor
            $a^{-4 } $''. The abscissa of this tip is therefore marked by the
            symbol $\mu_U$ (with a $\mu$ written by the curve progarm).
            Because the Einstein Hilbert action has a dimension 2 different
            behavior from the just counting plaquettes, it is the abscissa of
            the tip of the parabola, which had an $a^{-6}$ weighting relative
            to the ``original'', which means the effective lattice link-size
            for the extraction of the Planck scale $E_{Pl}$ energy. One shall
            note from figure or the trivial algebra that denoting the
            abscissa for the peak of the ``original'' by $\mu_0$ then the
            pushing of this
            tip energy scale by the two different linear extra terms
            in the logarithm by the $a^{-4}$ and $a^{-6}$ respectively makes
            displacements in the dominant (energy)scale by terms in the
            logarithm being in the ration 4 : 6 = 2 : 3. This means the
            prediction
            \begin{eqnarray}
              \frac{\ln(\frac{\mu_U}{\mu_0})}{\ln(\frac{E_{Pl}}{\mu_0})}
              &=& \frac{4}{6}=\frac{2}{3}.
              \end{eqnarray}
            But we have to guess e.g $\mu_0=M_Z$ or $\mu_0=m_t$ to use this.

            \subsection{Variants of the Relation Planck Scale to
              ``Our Unified''}

            In fact the scales $\mu_0$ and also the ``Planck scale'' do
            not come in precisely from our physics and are at best order of
            magnitudes wise determined. The $\mu_0$ scale should be
            where the effective number of families is having maximum,
            but honestly this effective number of families is 3 from the
            top-mass and up to infinity? So we only know $\mu_0 \ge m_t$.
            And for the Planck scale we shall in fact claim that it would
            be expected that the reduced Planck scale (including the often
            associated $8\pi$ to $G$ before using dimensional arguments
            to construct an energy scale) is actually more reasonable to use.
            
            \subsection{How Well Agrees Our Relation?}
            
            In the table \ref{tawgr} we therefore combine   our fit above
            obtained value of the ``our unification scale''
            $\mu_U$  = $5.1*10^{13}$ GeV with some reasonable suggestions
            for the two less welldefined scales $\mu_0$ and the
            ``Planck scale''.

            \begin{table}[!]
 \caption{\label{tawgr}
                 {\bf Table with $\mu_U$  = $5.1*10^{13}$ GeV of}
            $\frac{\ln \hbox{``gravity scale''}}{\ln \hbox{``unified scale''}}
             $=$ \frac{\ln(\frac{E_{Pl\;or \; Plred}}{\mu_0})}
                   {\ln(\frac{\mu_U}{\mu_0})} $
                  }
            \begin{tabular}{|c|c|c|c|}
          \hline
          Name & $\mu_0$&$E_{Pl}$& $E_{Pl\; red}$\\
          &  &=$1.22*10^{19}GeV$&= $2.34*10^{18}GeV$\\
          \hline
          $Z^0$ mass &$M_Z$ &1.4579&1.3968 \\
          &=91.1876GeV&&\\
          \hline
          Av. fermion mass&$m_{av. fermions}$ &&\\
          &=163 MeV&1.3711&1.3216\\
          \hline
          Top quark&$m_t$&&\\
          &=172.52 GeV&1.4689&1.4079\\
          \hline
          Fitted $\mu_0$&$\mu_{0\; best}$&&\\
          &=24.231 TeV& 1.5769&1.5 (exact)\\
          \hline
            \end{tabular}
           
            \end{table}

            \subsection{Fitted $\mu_0$}
            
        Alternative to just guessing on good ideas of what our scale $\mu_0$
        at which the density of the size of the scale is maximal counting with
        its own link length as unit we can simply fit, what we would like this
       scale to be and then if possible build up a story of, what it should be
        of that order. Such a fitting of the scale $\mu_0$ would simply mean,
        that we solve the equation of our prediction, say
        \begin{eqnarray}
        \ln(\frac{E_{Pl\; red}}{\mu_0}) &=& \frac{3}{2}*\ln(\frac{\mu_U}{\mu_0})\\
        \hbox{formally giving:}\frac{1}{2}\ln(\mu_0) &=&\frac{3}{2}\ln(\mu_U)
        - \ln(E_{Pl  \; red})\\
        &=& 31.563*3/2 -42.2967\hbox{(using GeV)}\\
        &=& 5.048\\
        \hbox{So: }\ln(\mu_0) &=& 2*5.048 \hbox{(in GeV used)}\\
        &=& 10.095 \hbox{(with GeV)}\\
        \Rightarrow \mu_0 &=& 24.23 TeV.
        \end{eqnarray}

        The choice of $\mu_0$ that would make prediction perfect was
        as seen from the table and our calculation $24 TeV$, which is
        higher than the top quark mass only by a factor
        $\frac{24.23}{172.25 GeV} = 7.11$. Very speculatively one could
        attempt to construct some fitting of the density as a function of
        the scale fermions still findable as effectively massless at the
        scales considered. Above the top mass of course the effective
        number of massless Fermions at the scale  correspond to 3 families,
        but as one asks below the top qaurk mass there is a 1/3 or 1/4 of a
        family missing, and one could claim that just below the top mass
        we have some 3-1/4 or 3-1/3 families left, and then crudely estimate in
        this spirit, if one could fit the (non-integer) number of families
        to a function reaching a bit above the top-mass a maximum of 3
        families. Then this maximum in a curve taken as a function of the
        logartihm of the scale would have its maximum with value 3 families
        very close to our wanted $\mu_0=24.23 TeV.$ (most wellcome
        to make gravity scale match our model).

        Such a very crude and somewhat arbitrary extrapolation might be this:
        
        Firts have in mind that the derivative of the number of
        ``effectively massless'' fermion families as function of the logarithm
        of the scale is given by the ``density of the fermions with mass
        at that scale''=``density of species''.
        
        On a logarithmic mass scale there is a ``density of species'' of 1 over
        a scale dististance $\ln \frac{m_t}{m_b} = 6.02$, meaning density
        = 0.166 per e-factor. The center of this interval is the geometric mean
        of 172.25 GeV and 4.180 GeV, which is 26.83 GeV. Next take the interval
        beteen the b-mass and the s-mass which has length
        $\ln( \frac{4.180}{0.095})= 3.784$ and contains two species quarks and
        3 species fermions, if we include the $\tau$-lepton. This means for this
        region around $\sqrt{4.180*0.095} GeV$ = 0.630 GeV we have the density
        of quark species $2/3.784 = 0.5285$. This density is 3.184 times
        bigger than in the interval between t and b. If the deviation from the
        maximal number 3 of the number of families at the  scale being
        seen as massless were varying with the logarithm of the scale
        quadratically counted out from some center value of the scale,
        then the slope of it, being in fact the ``density of species'' would
        vary linearly, and we should just extrapolate the density to the
        point, where it passes zero to find the maximum point for the
        formal number of massless families. In logarithmthe distance
        between the two points we considered is $\ln(\frac{26.83}{0.630})
        = 3.75$,and the linear extrapolation leads to the zero for the slope
        point displaced upward from the 26.83 by the exponential of
        3.75/(3.184-1)= 1.717, and that gives 149.4 GeV. It is
        actually 
        very close to the top-mass. Now we see, that if we take the
        difference between
        the two  possibilities we mention in the table for the Planck scale
        as an estimate of the uncertainty of only order of magnitude numbers,
        then this uncertainty for the ratio given in the table is of order
        0.06. But the best of the points actually for the top mass
        as $\mu_0$ deviates only by 0.03 from the predicted 1.5, so we must say
        that we should take it as agreement within  expected uncertainty.

        In any case we have shown how a fluctuating lattice size speculatively
        can solve our problem, that the unification scale is quite
        different from the Planck energy scale,in spite of that we
        want a common lattice to  describe them both.

        \section{Conclusion}
\label{ps12}
We have succeeded in constructing a lattice model picture, in which
we fit the three fine structure constants in the Standard Model by
{\bf three parameters, which are with limited accuracy predicted by various
assumptions of the model}. What we consider the most important is,
that we suggest, that the way with smallest representation used as the
link variables for the Standard Model {\bf group} - understood as the
global group structure $S(U(2)\times U(3))$ in the O'Raifeartaigh sense
\cite{OR} and not only
the Lie algebra - is in fact links, that also could have been an $SU(5)$
representation, and therefore the model obtains an {\bf approximate }
$SU(5)$-symmetry, when we imposed the usual trace-action. We then took it,
that this first approximation
$SU(5)$-symmetry of the classically treated simple trace action
was broken by quantum fluctuations, which  are of course only present
for those fluctuations, which are true standard model group degrees of freedom,
while the degrees of freedom which are only in $SU(5)$, but not in the
Standard model group, of course do not contribute quantum corrections to
correct the finestructure constants in our model, wherein they do not exist.
It is this quantum correction breaking the $SU(5)$ symmetry (The $SU(5)$
relation between the couplings is  only valid  in the classical
approximation) that brings the deviations from
$SU(5)$ GUT theories {\bf without help from additions as susy}, and indeed we
``predict''
in our model not only ratios of the shifts caused by the quantum fluctuation
for the three different standard modelinverse fine structure constants,
but also {\bf the absolute size of the corrections}. So even, if we used say
the ratio of the corrections to fit the pseudo-unified scale, or let us
say ``our unification scale'' $\mu_U$, then it is still a prediction, that
we know the {\bf size of the correction} from precise $SU(5)$ unification. This
prediction - it must be admitted though - contains a factor 3 being the
number of families. Really it is the number of parallel lattices supposed
to exist in Nature, that is 3, so the connection to the number of families
is, that there would be by assumption one layer (one of the Wilson lattices
lying in parallel) for each family of fermions. (Each family its own
``layer''.)

The success of this predicting the {\bf deviation} from GUT by quantum
corrections fits actually the to experiment fitted fine structure constants
at say the $M_Z$ ($Z^0$ -mass scale) {\bf within uncertainties}! And this
is quite
remarkable, because these uncertainties for the three inverse finstructure
constants in the Standard Model are much smaller by a factor of the order
of 50 than the corrections due to the quantum fluctuations, we predicted.

It is  due to the high accuracy, with which the fine structure constants are
- now a days - known, that we can find so good agreement  compared to our
quantum corrections, because these corrections are indeed about 10 times
smaller than the typical inverse fine structure constant, which is of order
40, while our correction are of the order of 1 times the important
``unit'' for our corrections $3 *\pi/2 = 4.7124$. In fact
we predict e.g. the difference between the inverse fine structure
constants at the ``our unification scale '' ($\mu_U$) such as 
        \begin{eqnarray}
          1/\alpha_2(\mu_u) - 1\alpha_3(\mu_u) & ``predicted''& 3*\frac{\pi}{2}
          = 4.7124\\
          \hbox{turned out: } 1/\alpha_2(\mu_u) - 1\alpha_3(\mu_u)&`` fitted''
          & 4.62.
          \end{eqnarray}
        and the uncertainty in these inverse fine structure such as
        e.g. the $1/\alpha_3$ is $\pm 0.05$, so the deviation
        of $0.09$ is only $1.8$ s. d.(s.d.= standard deviations), and if
        we count two similar
        numbers the estimated uncertainty would be $\pm \sqrt{2}*0.05$
        = $\pm 0.07$ and we would have 1.3 s.d. Our deviation and uncertainty
        are of the order of a factor 52 smaller than the quantity of deviation
        4.62, which we found!

        It would in itself be interesting just to leave the
        two further parameters, namely the unified coupling - for the $SU(5)$
        - and the scale of this approximate unification, because we would
        even then have an interesting relation between the fine structure
        constants.
        
        \subsection{The Further Two Parameters}
        
        But we have also formally manged to find assumptions, so that
        these two further parameters are fitted within the now somewhat
        smaller accuracies:
        \begin{itemize}
        \item{The Unified Coupling as Critical Coupling}
          
        We mannaged to be allowed to claim, that the unified coupling
        is indeed the critical coupling for the non-exitent $SU(5)$ in our
        model. So in a way there is the little worry with this prediction:
        that it is for the $SU(5)$ lattice gauge thoery, we use the critical
        coupling, but this $SU(5)$ theory is not truly present in our model.
        One should possibly replace the $SU(5)$ critical couplng by one for
        a modified $SU(5)$ with the degrees of freedom cut down to those
        of the Standard Model -like it is in our model - but such a correction
        would make the critical coupling be stronger(i.e. lower
        $1/\alpha_{5\;crit}$), and that would make the fitting with this
        critical coupling being the unified one worse prediction.
        So after such improvement
        our unified coupling prediction would not work so well, if this was
        all we did. But if one starts from a standard model group critical
        coupling, then one should not make the quantumcorrections as if it
        were a full $SU(5)$. When we also correct the quantum correction to
        be for the Standard Model Group, then it actually seems to agree better.

      \item{Relation of the Unified Scale to the Planck Scale}

        Our story behind our formally within errors relating in our model
        our unified scale - at which our corrections are to be applied -
to the Planck scale
        may be a bit too much made up with guesses to be truly convincing.
        Thus this part of the work should rather than being  an
        attempt to
        find a third predicted parameter, namely the unification scale
        - what it though also is -, be taken as a needed story for rescuing
        our model against a severe problem: Our unification scale $\mu_U$
        should as the lattice scale be the fundamental scale in our model.
        But that is not so good, because this ``unification energy scale'' is
        much lower than the presumably fundamental scale of gravity,the
        Planck energy scale?

        \end{itemize}

        \subsection{Problem with Planck Scale in Our Model}

        (In the references \cite{RSR, BledRSR} we have after the
        present made progress on the fluctuating lattice, meant help
        on the problemof unified and Planck scale not matching)
        
     The problem with the Planck scale comes about like this:
     
     It is not surprising that this unified scale turns out, like in
     all GUT-theories, to be appreciably smaller than the Planck scale, and
     in our theory it is even compared to usual unification a bit small:
\begin{eqnarray}
       \mu_U &=& 5.13 *10^{13}GeV. 
\end{eqnarray}

However, the real problem is that we suggest to have a lattice that is taken
{\bf seriously to exist in Nature,} and we would seemingly loose  ordinary
continuum manifold physics for smaller distances than $1/\mu_U$ and
the seemingly approximate well working generel relativity taken classically
at such scales, would be already to be considered as a quantum gravity,
and in addition we would find it a priori non-attracktive to have several
(two) fundamental scales($\mu_U$ and the Planck energy scale).

\subsection{Gravity Has to Be ``Weak'' on Fundamental Scale}

     This may bring us some message about gravity: We have to invent a
     story of the kind, that gravity is for some reason very weak compared
     to the fundamental scale expectation. Our above described model namely
     has as its philosophy, that the unified scale - which remains low compared
     to the Planck scale in energy - is to be the `` fundamental scale''!
     You might speculatively think about, that the $g^{\mu\nu}$ (with upper
     indices) has appeared as kind of spontaneous breaking of e.g.
     diffeomorphism symmetry, and and thus has a chance to be small
     (often one finds relatively small spontaneously breaking fields,
     otherwise it would not be so common with low temperature super
     conductivity, that it was a big sensation to find high temperature
     super conductivity). If this $g^{\mu\nu}$ is small compared the
     our fundamental lattice, then compared to this lattice the $g_{\mu\nu}$
     with lower indices will be large and thus the length say of a lattice link
     would be big. This bigness would be bigness compared to the Planck
     constant and so getting $g^{\mu\nu}$ by some spontaneous breaking story
     would help bringing about the lack of coincidence of our fundamental scale
     with the Planck one\cite{Bled2023}.

     Although this idea of having $g^{\mu\nu}$ representing a spontaneous
     symmetry break down and being ``small'' for that reason, seems
     attracktive to me, we shall in this article rather seek to solve
     the problem with the Planck scale being different from ``the our unified
     one'' $\mu_U$ by the idea of fluctuating lattice link size described in
     next subsection.
     
     \subsection{Fluctuating Lattice Scale}
     
     A priori it seems somewhat embarrasing, that our theory taken seriously
     wants a fundamental scale with lattice already at the approximately
     unification scale $5*10^{13}GeV$, while we a priori would expect
     the fundamenntal scale at the Planck scale, especially for the gravity
     itself, when we even seek to uphold a principle of critical coupling
     constants
     .
     If a lattice gravity should have in one sense or another a critical
     coupling, then the lattice should be of the Planck scale lattice constant
     roughly. The speculation solution, that almost has to be needed is, that
     of the in scale fluctuating lattice like this or something similar:

     At around the ``unifying scale'' the gravitational fields must
     behave classically to a very good approximation, except though that a gauge
     degree of freedom  would tend to fluctuate infinitely (actually Ninomiya
     F{\"o}rster and myself\cite{FNN} even would let such strong quantum
     fluctuations
     be the reason for the exact gauge symmetry.) because there is lacking
     terms in the Lagrangian, that can keep the gauge to a fixed one, except the
     by hand put in gauge fixing terms, but they are of course not physical.

     This then means, that we must think in a gravity containing theory as the
     lattice fluctuating being dense with small lattice constant somewhere
     in the Riemann space-time and large somewhere else. In that case we must
     imagine that the ``observed'' lattice scale (as for our model
     the $5.13*10^{13} GeV$) will be some appropriate average over a highly
     fluctuating lattice constant size. We would expect the local lattice scale
     to fluctuate with a distribution that would be an approximately flat
     distribution in the logarithme of the lattice constant, because the
     diffeomorphism group contains scalings and the Haar measure for a pure
     scaling symmetry subgroup would suggest a smooth in logarithm
     distribtution. But now, while the averaging of the Yang Mills Lagrangiam
     over a distribution of scales with a smooth distribution in the logarithm
     would be weighted in slowly varying way, the gravity action, the
     Einstein Hilbert one varies with a power law with the scale of the
     lattice, if you, as we had success with, assumed a critial coupling.
     This would then lead to that the average size of the lattice link or
     plaquette structures contributing dominantly to gravity action would be
     much smaller than the ones contributing to the Yang Mills fields action.

     This could suggest a mechanism for the seeming fundamental scale
     (= lattice constant size scale) for gravity would be much higher in
     energy than for the Yang Mills theories.

     A fluctuating lattice might provide a natural explanation for
     the much smaller Planck length than length scale at the Yang Mill.
     
     \subsection{Baryon Non-Conservation ?}
     
     Our theory is in danger of inheriting baryon decay in analogy to the
     usual $SU(5)$ grand unification theories, but at least the
     gauge particles in the $SU(5)$ theory which are not in one of the
     standard modelgroups also are supposed not to exist in our scheme,
     so the obvious diagram with an exchange of such an $SU(5)$ gauge particle
     is missing in our model. Actually it is in our model some four fermion
     interaction, that could give the baryon violation, but such an interaction
     would have a dimension similar to that of the Einstein Hilbert action,
     and thus the interaction of such a type violating baryon number
     conservation would be suppressed as a term in Lagrangian of
     high order with Planck energy as the energy unit.
     At least that is, what happens in our model, just using our cut off
     scheme as we did with gravity (fluctuating lattice scale). Whether our
     Gaussian in log weighting
     can be assumed sufficiently consistently to suppress the baryonnumber
     violation suffiently to  cope with bounds on proton decay may deserve
     study in a
     later work, but at first it looks like working and giving sufficient
     suppression.
     .

When we in this way have no other breaking of the baryon number than via the
instantons present in the Standard Model, it means that our model points
towards baryon assymetry should be caused via a lepton assymetry.
We namely only get the instanton baryon number variation
to wash away any previous baryon number assymetry unless there is
a lepton number assymmetry. But this pointing to  the baryon assymetry
comming from a lepton assymetry is not an absolute must, because the baryon
number concervation is still only an accidental symmetry in our model; we only
got rid of the for true SU(5) so embarrasing violation of baryon conservation.

That we have a see-saw scale much lower than the Planck scale may also
encourage us that even baryon violation could be there too, if it could help
the assymetry without making decays. But we do favor leptonassymetry as the
mechanism.
     
     \subsection{Is Approximate Scale Invariance a Dirty Assumption?}

     Of course, when we claim, what we in this fluctuating lattice model
     claim, that we have on the one hand an at least approximate scale
     invariance;
     but nevertheless, that this symmetry is broken so much, that the size
     distribution of the numbers of lattice links, or of lattice plaquettes,
     has maximum at some finite scales - even in order of magnitude -
     depending on the exact weighting, this sounds a bit dangerous, and  can
     only be true approximately. 
     The meaning is e.g., that if you include some extra power of the
     (inverse) link
     length $(1/a)^n$, it can shift maximum in the size distribution
     from e.g. ``the 0ur unifying scale'' to the Planck scale. It also involves
     some physical effect or principle, that performs the needed very strong
     suppression of links or plaquettes being stronger and stronger the
     smaller the link or plaquette.

     At first it looks like breaking reparametrization invariance in general
     relativity, does not sound nice. But we must postpone this problem
     just having now admitted, that there is a problem, that would need more
     detailed modelling, and that most likely such improved models would be
     too complicated to be believable.

\subsection{Our Progress Compared to Our Earlier Works}

One way of looking at the progress of the present work is to think of it as
an updated version of the work by Don Bennett and myself\cite{Don137}, which
seeks to get all the three fine structure constant from criticallity at
Planck scale and the antiGUT type of model with the gauge group being
a cross product of 3 isomorphic Standard Model groups. But in the old works
we had to help by extra assumptions the $U(1)$ fine structure constant.
In the present article this helping the $U(1)$ has been replaced by the
approximate $SU(5)$, so that it seems more natural, and not so specially just
making some story for $U(1)$ alone.
\subsection{Outlook}
     \subsubsection{The Dream of Exact Formula for $\alpha_{EM}$}
     
     Of course behind such fittings of finestructure constants is the holy
     gral dream of finding the mathematical formula for the
     (electrodynamics) fine structure
     constant, because that is so well known - many decimals - that it contains
     so much infromation\cite{Rugh} that one could hope to justify a theory
     to be correct, if it fitted the fine structure constant in a sufficiently
     simple way (with the many decimals). A work like the present would
     suggest restrictions on the form
     of the formula for the fine structure constant, and thereby make an a bit
     more complicated formula be acceptable as convincingly right provided it
     were of the right form.

     But to make a formula without from phenomenology included expressions
     possible we would of course need to have the Higgs and the fermion masses
     connected, and for the time being the usual philosophy is, that the
     Higgs scale is a pure mystery and, that it needs a solution of the hirarchy
     problem to be possible at all. Some different philosophy e.g.
     a coupling of the weak scale or Higgs scale to the development
     of the renorm group (for e.g. the top quark mass) is needed, one example
     is our \cite{Coincidensies,Rudjer} applying the complex action
     theory also in \cite{Relation}.
     
     \subsubsection{Could the See-Saw Scale Be Identified with Our Unifcation
       Scale?}
     It is characteristic of the our unified scale $\mu_U$ for the only
     approximate that it is a bit to the low side in energy to even unified
     scales in other models (especially if it is models with susy), and
     further it is the spirit of our model that since our unification
     scale is a lattice scale - or some dominating average in a fluctuating
     lattice link size -. It is only $5.13*10^{13} GeV$. So it puts us in
     the direction of asking if the see-saw mass scale could be the same
     as our unification scale?

     The neatrino mass square differences are for the atmospheric
     netrino mass square difference and the solar one
     \begin{eqnarray}
       \Delta m_A^2 &\approx& 1.4 *10^{-3}eV^2 \hbox{ to }3.3*10^{-3} eV^2\\
       \Delta m_{sol}^2 &\approx & 7.3*10^{-5}eV^2 \hbox{ to } 9.1*10^{-5} eV^2 
     \end{eqnarray}
     indicating masses of the order of magnitudes
     $ (4 \hbox{ to } 5)*10^{-2} eV $ and $3*10^{-3}eV$. With say
     a typical charged fermion mass in the Standard Model being of
     mass $1GeV$, you would expect by dimensional arguments a see saw neutrino
     mass of the order
     \begin{eqnarray}
       \hbox{``see saw scale''} & \approx & \frac{(1GeV)^2}{10^{-2} eV}\\
       &=& 10^{11} GeV\\
       \hbox{Not so far from our} \mu_U &=& 5.13*10^{13}GeV. 
       \end{eqnarray}
     If we take it that the spread in the charged fermion masses from the
     electron mass $0.5*10^{-3} GeV$ and the top quark $174 GeV$ implies
     that our typical charged fermion mass shall be considered to  have
     2 to 3 orders of magnitude uncertaity, implying by the squaring
     in going to the see-saw mass a doubling in the numbers of orders
     of magnitude, then the see-saw scale is
     \begin{eqnarray}
       \hbox{``see saw scale''} & \approx & 
       10^{11} GeV*10^{\pm 5}\\
       \hbox{having inside errors }\mu_U&=& 5.13*10^{13} GeV.
       \end{eqnarray}
     So if we believe in a lattice already at the $5.13*10^{13} GeV$,
     we can look for replacement of the see-saw neutrinos by some
     lattice effects.
     \subsubsection{Small Hierarchy by the Charges from
       $G_{SMG}\times \cdots \times G_{SMG}$   }
     If our model were right one would look for understanding the charged
     fermion masses along the lines of our old work with Yasutaka Takanishi
     and Colin Froggatt \cite{Takanishi}, while the neutrino oscillations
     would be related to the lattice of effective lattice scale only
     $5.13*10^{13} GeV$.

     \subsubsection{Modification of Accurate Standard Model Results}
\label{threferees2}
The idea of the fluctuating lattice put forward above and the hypotesis
that the lattice is truly existing means that ideal Standard Model perturbative
calcuæations studied with high accuracy, such as e.g. the famous anomalous
magnetic moment calculations, would be modified because they should be
performed with truly exisitng lattice, which only has a finite link-length.
In fact in the philosophy of the fluctuating lattice stricly speaking the
central value for the link distribution is about the top mass or better say
$10^4 GeV$, which means that corrections to ideal perturbation theory would
be much closer than if the most fundamental scale had been the Planck scale.

Also it would be very great of course, if one could experimentally somehow
explore the non-locality due to relatively seldom lattice links of
exceptionally large lengths. Indeed also such effects should be much more
accesible if the present fluctuating lattice was true, than say Planck scale
were the fundamental scale.

     \section*{Acknowledgement}
     I am thankfull to the Niels Bohr Institute for allowing me as emeritus,
     and then I am of course thankful for all the discussions during the earlier
     related works, although some of it  is now many years ago to
     Niels Brene, Don Bennett, Larisa Laperashvili, Ivica Picek and
     the important collaborators together with  Larisa were very close
     the same subject of combining Anti-GUT and GUT, D. A. Ryzhikh, C.R.Das...

     I
     remember,
     that Svend Erik Rugh was the first to tell me,
     that in fact the SU(5) GUT coupling was crittical (for SUSY-GUT) the
     unified coupling inverted is down to $1/\alpha_5(10^{16} GeV) \sim 25$
     and closer to just one of the above mentioned critical (\ref{cr}), $15.3
     \pm 7.5 \% $, but without the factor 3. I thenk the referees for small
     additions, e.g subsections \ref{threferee1} and \ref{threferees2}. 
     


\end{document}